\documentclass[twocolumn,superscriptaddress,amsmath,amssymb,floats,notitlepage]{revtex4-1}
\usepackage[latin9]{inputenc}
\usepackage{amsmath}
\usepackage{graphicx}
\usepackage{comment}
\usepackage{tikz}
\usepackage{float}

\makeatletter

\@ifundefined{textcolor}{}
{%
 \definecolor{BLACK}{gray}{0}
 \definecolor{WHITE}{gray}{1}
 \definecolor{RED}{rgb}{1,0,0}
 \definecolor{GREEN}{rgb}{0,1,0}
 \definecolor{BLUE}{rgb}{0,0,1}
 \definecolor{CYAN}{cmyk}{1,0,0,0}
 \definecolor{MAGENTA}{cmyk}{0,1,0,0}
 \definecolor{YELLOW}{cmyk}{0,0,1,0}
}

\definecolor{ssdw}{RGB}{145,197,226}
\definecolor{csdw}{RGB}{242,119,98}
\definecolor{svc}{RGB}{165,217,137}
\definecolor{ssdwpcsdw}{RGB}{100,129,189}
\definecolor{ssdwpsvc}{RGB}{0,166,81}
\definecolor{csdwpsvc}{RGB}{201,143,176}
\definecolor{ssdwpcsdwpsvc}{RGB}{255,196,102}

\usepackage{epstopdf}
\usepackage{color}
\usepackage{array}
\usepackage{mathrsfs}
\usepackage[normalem]{ulem}
\usepackage{dcolumn}
\usepackage{bm}

\newcolumntype{L}[1]{>{\raggedright\let\newline\\\arraybackslash\hspace{0pt}}m{#1}}
\newcolumntype{C}[1]{>{\centering\let\newline\\\arraybackslash\hspace{0pt}}m{#1}}
\newcolumntype{R}[1]{>{\raggedleft\let\newline\\\arraybackslash\hspace{0pt}}m{#1}}

\graphicspath{{Impurities and Gaps/Plots/}}

\newcommand{\mbf}[1]{\mathbf{#1}}

\usepackage{tikz}
\usetikzlibrary{calc,intersections}
\usetikzlibrary{shadings}
\usepgflibrary{shadings}

\makeatother

\begin{document}


\title{Magnetic phase diagram of the iron pnictides in the presence of spin-orbit coupling: \\ Frustration between $C_2$ and $C_4$ magnetic phases}

\author{Morten H. Christensen}
\email{mchrist@umn.edu}
\affiliation{School of Physics and Astronomy, University of Minnesota, Minneapolis, Minnesota 55455, USA}

\author{Peter P. Orth}
\affiliation{Department of Physics and Astronomy, Iowa State University, Ames, Iowa 50011, USA}
\affiliation{Ames Laboratory, U.S. DOE, Iowa State University, Ames, Iowa 50011, USA}

\author{Brian M. Andersen}
\affiliation{Niels Bohr Institute, University of Copenhagen, Juliane Maries Vej 30, DK-2100, Denmark}

\author{Rafael M. Fernandes}
\affiliation{School of Physics and Astronomy, University of Minnesota, Minneapolis, Minnesota 55455, USA}

\date{\today}

\begin{abstract}
We investigate the impact of spin anisotropic interactions, promoted by spin-orbit coupling, on the magnetic phase diagram of the iron-based superconductors. Three distinct magnetic phases with Bragg peaks at $(\pi,0)$ and $(0,\pi)$ are possible in these systems: one $C_2$ (i.e. orthorhombic) symmetric stripe magnetic phase and two $C_4$ (i.e. tetragonal) symmetric magnetic phases. While the spin anisotropic interactions allow the magnetic moments to point in any direction in the $C_2$ phase, they restrict the possible moment orientations in the $C_4$ phases. As a result, an interesting scenario arises in which the spin anisotropic interactions favor a $C_2$ phase, but the other spin isotropic interactions favor a $C_4$ phase. We study this frustration via both mean-field and renormalization-group approaches. We find that, to lift this frustration, a rich magnetic landscape emerges well below the magnetic transition temperature, with novel $C_2$, $C_4$, and mixed $C_2$-$C_4$ phases. Near the putative magnetic quantum critical point, spin anisotropies promote a stable Gaussian fixed point in the renormalization-group flow, which is absent in the spin isotropic case, and is associated with a near-degeneracy between $C_2$ and $C_4$ phases. We argue that this frustration is the reason why most $C_4$ phases in the iron pnictides only appear inside the $C_2$ phase, and discuss additional manifestations of this frustration in the phase diagrams of these materials.
\end{abstract}
\maketitle

\section{Introduction}

The phase diagrams of the iron pnictide superconductors display a rich structure, exhibiting, in addition to superconductivity, a multitude of magnetic phases~\cite{johnston10,wen11,greene_review,avci14a,meier17,wang16}. Elucidating the nature and origin of these magnetic phases constitutes an important part of understanding the origin of unconventional superconductivity in the iron pnictides~\cite{basov_review,chubukov_review}. Similar to other unconventional superconductors, the superconducting dome in the pnictides is centered around the end of a magnetic dome~\cite{scalapino12,dai_review}. However, in contrast to cuprates and heavy fermion compounds, which primarily exhibit N{\'e}el antiferromagnetism, the pnictides are dominated by orthorhombic stripe spin-density wave (SSDW) magnetic order, where the magnetic moments are anti-parallel only along one Fe-Fe direction, leading to the breaking of tetragonal symmetry~\cite{fernandes_review,bohmer_review}. Nonetheless, recent experiments~\cite{avci14a,wang16,hassinger,allred16a,bohmer15a,wasser15,khalyavin14,mallettb} have revealed the appearance of tetragonal magnetic order as magnetism is suppressed by doping or pressure.

The crystal structure of the pnictides in the paramagnetic phase is tetragonal and inelastic neutron scattering experiments reveal peaks at $\mbf{Q}_{1,2}=(\pi,0),(0,\pi)$ in the 1Fe/unit cell. This motivates considering two magnetic order parameters, $\mbf{M}_{1,2}$, with ordering vectors $\mbf{Q}_{1,2}$ related by $C_4$ symmetry. The condensation of only $\mbf{M}_1$ or $\mbf{M}_2$ leads to the SSDW phase mentioned above, and the choice of either $\mbf{Q}_1$ or $\mbf{Q}_2$ implies the breaking of tetragonal symmetry, leading to a preemptive or simultaneous nematic transition~\cite{fernandes12,avci12,kim11,rotundu11,kasahara12,zhou13}. Experimentally, the magnetic moments in the parent compounds are observed to lie in-plane, parallel to the ordering vector, i.e. $\mbf{Q}_i \parallel \mbf{M}_i$.

On the other hand, the possibility that both order parameters condense simultaneously leads to two additional options for the magnetic order~\cite{lorenzana08,wang14,gastiasoro15,christensen15,christensen17,
fernandes16,eremin,giovannetti,wang15,vandenbrink17,fernandes16}. Collectively, these are referred to as $C_4$ magnetic orders as they leave the tetragonal symmetry of the lattice intact. One, with $\mbf{M}_1 \parallel \mbf{M}_2$ and $|\mbf{M}_1|=|\mbf{M}_2|$, is the charge-spin density wave (CSDW) phase, for which the magnetization is non-uniform and vanishes on half the Fe sites~\cite{allred16a}. This induces a secondary checkerboard charge order~\cite{lorenzana08}, thus motivating the name. The other has $\mbf{M}_1 \perp \mbf{M}_2$ (also with $|\mbf{M}_1|=|\mbf{M}_2|$) and is dubbed the spin-vortex crystal (SVC) phase due to the vortex-like structures arising in the real-space magnetization profiles~\cite{fernandes16}. The $C_4$ magnetic orders have been observed to appear with hole-doping or pressure in a diverse range of materials such as Ba$_{1-x}$Na$_x$Fe$_2$As$_2$~\cite{avci14a,wang16}, Ba$_{1-x}$K$_x$Fe$_2$As$_2$~\cite{hassinger,bohmer15a,allred15a,mallettb}, Sr$_{1-x}$Na$_x$Fe$_2$As$_2$~\cite{allred16a,taddei16a}, Ca$_{1-x}$Na$_x$Fe$_2$As$_2$~\cite{taddei17a}, FeSe~\cite{bohmer18}, and Ni- and Co-doped CaKFe$_4$As$_4$~\cite{meier17}. Determining which type of $C_4$ magnetic order (CSDW or SVC) is present in these systems is experimentally challenging. However, in Sr$_{1-x}$Na$_x$Fe$_2$As$_2$, a M{\"o}ssbauer study demonstrated the presence of a CSDW phase with out-of-plane moments~\cite{allred16a}. The transition from orthorhombic to tetragonal magnetic orders occurs close to the edge of the magnetic dome, in the vicinity of a putative quantum phase transition from the paramagnetic to the magnetic state.
On the other hand, in Co- or Ni-doped CaKFe$_4$As$_4$ a combination of M{\"o}ssbauer and nuclear magnetic resonance (NMR) measurements showed a SVC phase with in-plane moments oriented $45^{\circ}$ to the Fe-Fe axis~\cite{meier17} (see Fig.~\ref{fig:mag_config}).

The fact that the moment direction seemingly depends on the type of magnetic order hints at the importance of spin anisotropy in these systems. Indeed, both polarized inelastic neutron scattering measurements~\cite{lipscombe10,qureshi12,wang13,song13,steffens13} and NMR~\cite{li11,hirano12,curro17} indicate the presence of substantial spin anisotropy in the pnictides. As discussed previously~\cite{li11,christensen15,dai16}, such spin anisotropy can be naturally accounted for by the sizeable spin-orbit coupling (SOC) observed by angle-resolved photoemission spectroscopy in these systems~\cite{borisenko16}.

In this paper we study the impact of spin anisotropies on a phenomenological description of the magnetic phase diagram, both at the mean-field level and beyond. The main results can be summarized as follows:
\begin{itemize}
	\item In the vicinity of the magnetic transition temperature, the system can exhibit phases in which two or more of the original phases, SSDW, CSDW, and SVC, coexist at a microscopic level. This results in e.g. double-$\mbf{Q}$ phases which break C$_4$ symmetry. These are stabilized by SOC-induced spin anisotropic terms in the action. The presence of such spin anisotropic terms may cause frustration between the possible types of magnetic order. Frustration occurs when the magnetic ground state, which is determined by interactions (i.e. quartic coefficients of the action), becomes incompatible with the spin anisotropies imposed by SOC. Thus, the ground state obtained in the absence of spin anisotropy can be incompatible with the moment direction imposed by the SOC. Near the magnetic transition temperature the spin anisotropy due to SOC is dominant and the frustration is therefore lifted. These results are discussed in detail in Sec.~\ref{sec:mean_field}.
	\item Degeneracies between different magnetic orders emerge close to the magnetic quantum critical point (QCP) due to the SOC. In the FeSC, a putative QCP is found as magnetism is suppressed by doping, although it is typically hidden by the superconducting dome. We demonstrate this phenomenon using a renormalization group (RG) approach. The SOC-induced spin anisotropy is a relevant perturbation and leads to a drastic modification of the RG flows. The enhanced magnetic degeneracy appears due to the Gaussian fixed point being stable for a large range of system parameters. This is in contrast to the spin isotropic case, in which the Gaussian fixed point is unstable. The RG approach is discussed in Sec.~\ref{sec:rg_eqs} and a concise account of these results was given in Ref.~\onlinecite{christensen17b}.
\end{itemize}
In addition to the main results discussed in Secs.~\ref{sec:mean_field} and \ref{sec:rg_eqs}, in Sec.~\ref{sec:mag_free_energy} we introduce the model and provide further background, while in Sec.~\ref{sec:conclusions} we discuss the implications of our results. In Appendix~\ref{app:diagrams} we provide further details for the derivation of the RG flow equations.

\section{Phenomenological model for the magnetic phase diagram}\label{sec:mag_free_energy}

\begin{figure}
\centering
\includegraphics[width=0.9\columnwidth]{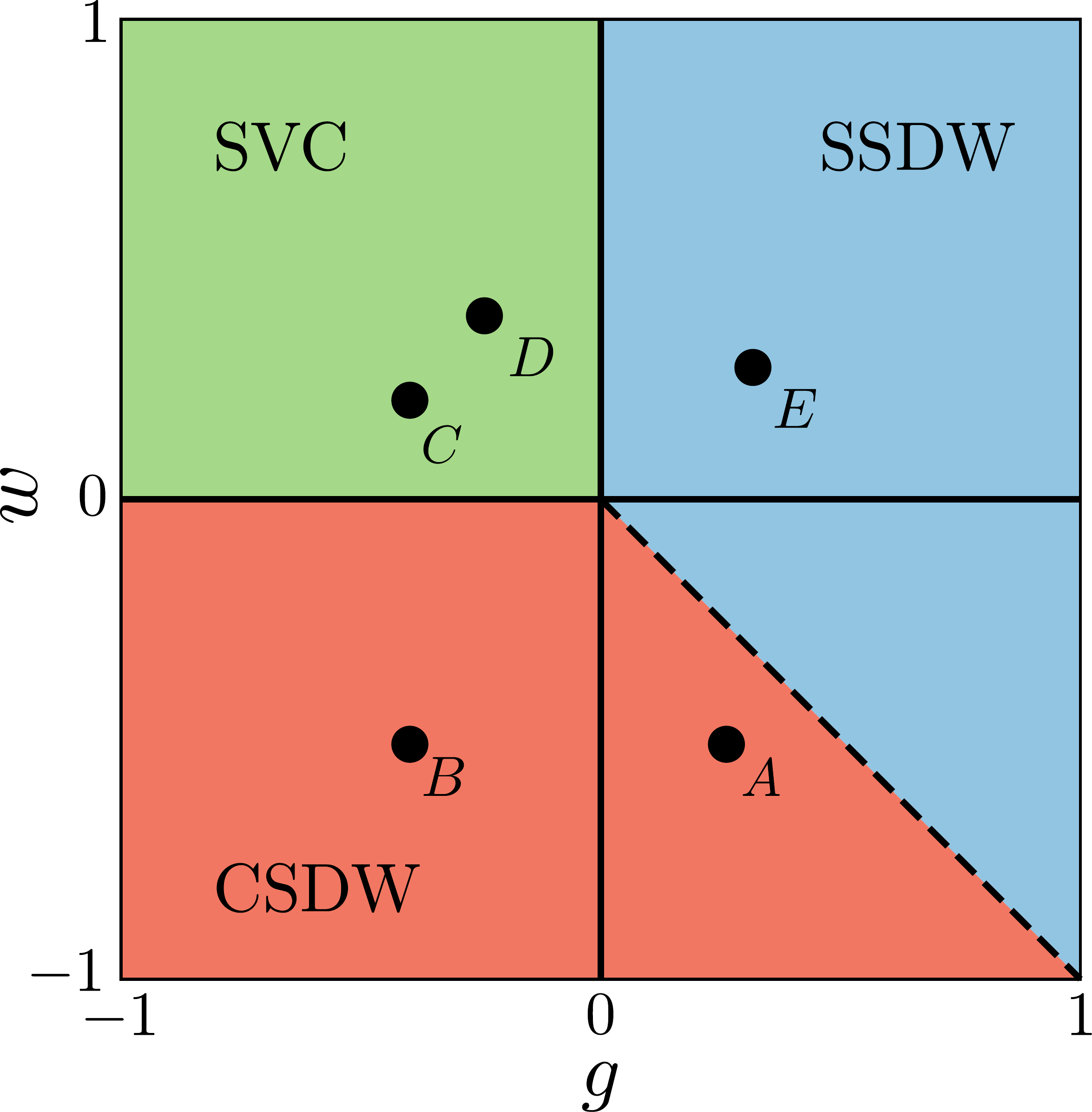}
\caption{\label{fig:mf_phase_diagram_w_points} (Color online) Mean-field phase diagram of the action in Eq.~(\ref{eq:action_1}) as a function of the two quartic coefficients $g$ and $w$. The boundedness of the free energy requires that $u>g$ in the SSDW phase, $u>-w$ in the CSDW phase, and $u>0$ in the SVC phase. The points $\{A,B,C,D,E\}$ indicate the parameter sets of $g$ and $w$ which will be discussed in detail in Sec.~\ref{sec:mean_field} within mean-field theory (see Figs.~\ref{fig:case_a}--\ref{fig:case_e}).}
\end{figure}

Many of the parent compounds of the iron pnictides are striped antiferromagnets with ordering vectors $\mbf{Q}_{1,2}=(\pi,0),(0,\pi)$ in the 1Fe/unit cell. Hence, in the ordered phase, the magnetic moment at each Fe-site is
\begin{eqnarray}
 	\mbf{m}(\mbf{R}) = \mbf{M}_1 \cos \mbf{Q}_1 \cdot \mbf{R} + \mbf{M}_2 \cos \mbf{Q}_2 \cdot \mbf{R}\,.
\end{eqnarray}
Prior to the formation of magnetic order the systems exhibit tetragonal symmetry. Together with time-reversal symmetry this restricts the form of the action, which can be written as~\cite{lorenzana08,eremin,giovannetti,wang14,wang15,fernandes16}
\begin{eqnarray}
	\mathcal{S}[\mbf{M}_1,\mbf{M}_2] &=& \frac{1}{2}\int_q r_0(q) \left(|\mbf{M}_1(\mbf{q})|^2 + |\mbf{M}_2(\mbf{q})|^2 \right) \nonumber \\
	&+& \frac{u}{2}\int_x \left( \mbf{M}^2_1(\mbf{x}) + \mbf{M}^2_2(\mbf{x})\right)^2 \nonumber \\
	 &-& \frac{g}{2} \int_x \left( \mbf{M}^2_1(\mbf{x}) - \mbf{M}^2_2(\mbf{x})\right)^2 \nonumber \\
	 &+& 2w\int_x \left(\mbf{M}_1(\mbf{x}) \cdot \mbf{M}_2(\mbf{x}) \right)^2\,, \label{eq:action_1}
\end{eqnarray}
where $q=(i\omega_n,\mbf{q})$ and $x=(\tau,\mbf{x})$ with the integrals $\int_{q}\equiv T\sum_{\omega_{n}}\int\frac{\mathrm{d}^{2}q}{(2\pi)^{2}}$ and $\int_{x}\equiv\int_{0}^{1/T}\mathrm{d}\tau\int \mathrm{d}^{2}\mbf{x}$. Here we consider a two-dimensional system, as the coupling between neighboring FeAs-layers is weak.
\begin{figure}
\centering
\includegraphics[width=0.95\columnwidth]{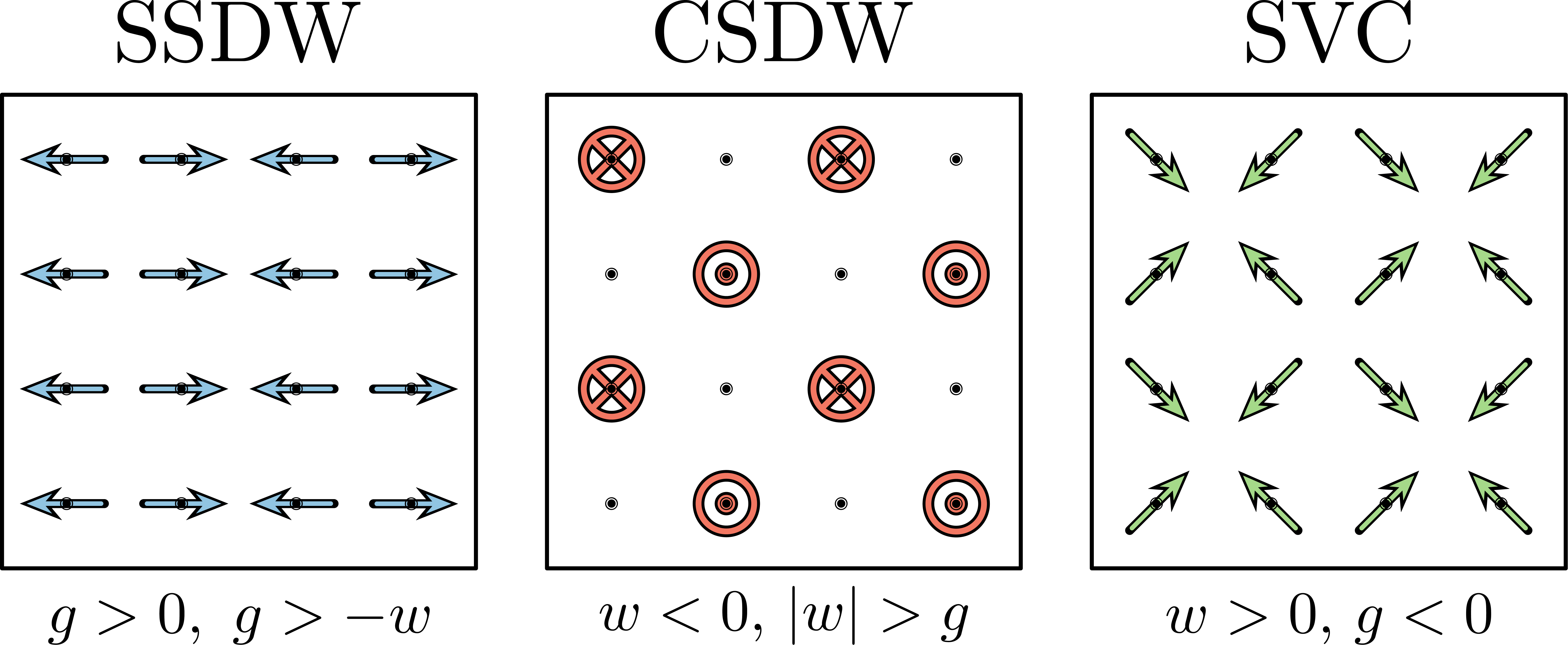}
\caption{\label{fig:mag_config} (Color online) Possible commensurate magnetic ground state configurations $\mbf{m}(\mbf{R})$ obtained from the free energy. Only Fe sites are shown. Note that, in the absence of spin anisotropy, the free energy only fixes the relative orientation of the magnetic moments. The directions were chosen to match those observed in experiments. Here, SSDW denotes the stripe spin density wave phase, CSDW the charge-spin density wave phase and SVC the spin-vortex crystal phase.}
\end{figure}
The quadratic coefficient, $r_{0}(q)=r_{0}+\mbf{q}^{2}+\gamma|\omega_{n}|$, is the bare inverse susceptibility with bosonic Matsubara frequency $\omega_n=2\pi n T$ and Landau damping parameter $\gamma$.  In the high-temperature classical limit, $r_0$ has the form
\begin{eqnarray}
	r_0=a(T-T_{\rm mag})\,,\label{eq:r_0}
\end{eqnarray}
where $a>0$ and $T_{\rm mag}$ is the mean-field magnetic transition temperature in the absence of SOC. In the $T=0$ case, $r_0$ tunes the distance to the mean-field QCP. Damping of the magnetic fluctuations in these metallic systems occurs via excitations of particle-hole pairs and is thus Ohmic and described by a dynamic critical exponent of $z=2$. The free energy is obtained through
\begin{eqnarray}
	\mathcal{F} &=& -T\ln \mathcal{Z}\,, \\
	\mathcal{Z} &=& \int \mathcal{D}\left[\mbf{M}_1,\mbf{M}_2\right]e^{-\mathcal{S}[\mbf{M}_1,\mbf{M}_2]}\,. \label{eq:part_func}
\end{eqnarray}
Encouraged by experimental results, we focus on homogeneous and commensurate phases. In this case three separate minima are possible. These correspond to the three magnetic states, SSDW, CSDW, and SVC, which are depicted in Fig.~\ref{fig:mag_config}. Here we review the parameter regimes in which each phase is found, along with the constraints imposed on $u$ to ensure a stable bounded free energy in each case~\cite{lorenzana08,wang14}.
\begin{itemize}
	\item The SSDW phase is selected for $g>0$ and $-w<g$. The free energy functional is bounded for $u>g$.
	\item The CSDW phase is selected for $g<|w|$ and $w<0$. The free energy functional is bounded for $u>-w$.
	\item The SVC phase is selected for $g<0$ and $w>0$. The free energy functional is bounded for $u>0$.
\end{itemize}
Other than having to fulfill the above stability requirements, $u$ does not play a role in determining the magnetic order. The leading instabilities can thus be described entirely in terms of $g$ and $w$, as seen in Fig.~\ref{fig:mf_phase_diagram_w_points}, in which it is assumed that the free energy is bounded.

In the absence of SOC, the O(3) spin rotational symmetry and the lattice symmetries are completely decoupled. The moment direction is thus independent of the lattice wavevectors $\mbf{Q}_i$ and is spontaneously chosen within the full O(3) manifold. In the presence of a finite SOC this is no longer the case and the O(3) symmetry is broken down already by the presence of the lattice, leading to spin anisotropy. In the pnictides, the staggering of the As atoms along with the observation that the moments are centered on the Fe sites lead to a specific anisotropy~\cite{cvetkovic13}. To leading order this can be written as~\cite{christensen15}
\begin{eqnarray}
	\delta\mathcal{F} &=& \frac{\alpha_1}{2} \left(M_{x,1}^2 + M_{y,2}^2 \right) \nonumber \\ &+& \frac{\alpha_2}{2} \left(M_{x,2}^2 + M_{y,1}^2 \right) \nonumber \\ &+& \frac{\alpha_3}{2} \left(M_{z,1}^2 + M_{z,2}^2 \right)\,.\label{eq:soc_free_energy}
\end{eqnarray}
Evidently, the relative values of the coefficients $\alpha_1$, $\alpha_2$ and $\alpha_3$ control the direction of the magnetic moments. 
For $\alpha_1 < \alpha_2,\alpha_3$ the moments are in-plane, along the direction of the ordering vector, $\mbf{M}_1 \parallel \hat{\mbf{x}}$ and $\mbf{M}_2 \parallel \hat{\mbf{y}}$. For $\alpha_2 < \alpha_1,\alpha_3$ the moments are also in-plane, however, they are perpendicular to their respective ordering vectors, i.e. $\mbf{M}_1 \parallel \hat{\mbf{y}}$ and $\mbf{M}_2 \parallel \hat{\mbf{x}}$. Finally, if $\alpha_3 < \alpha_1,\alpha_2$, the moments point out-of-plane, $\mbf{M}_i \parallel \hat{\mbf{z}}$. The coefficients $\alpha_i$ were calculated in Ref.~\onlinecite{christensen15} using a low-energy model based on a $\mbf{k}\cdot\mbf{p}$-expansion around the $\Gamma$, X, and Y points. They were shown to be proportional to the SOC strength and Hund's coupling, $\alpha_i \propto \lambda^2 J_{\rm H}$, while the ratios between the $\alpha$s depend on the bandstructure and vary with doping. A more recent treatment considered the appearance of spin anisotropic terms in realistic band structures with SOC~\cite{scherer17}. Here, however, we will treat them as phenomenological parameters and study their impact on the magnetic phase diagram.

The quartic terms of the free energy are also modified by the finite SOC. Such modifications are proportional to $\lambda^2$ as well, but within our mean-field approach it is well-justified to neglect SOC  anisotropies in the quartic coefficients. The reason is that close to $T_{\rm mag}$ it is the quadratic coefficients that select which order parameter components condense and anisotropies in the quartic coefficients only become relevant at much lower temperatures. At low temperatures, we fully take the resulting anisotropies into account within our RG approach, discussed in Sec.~\ref{sec:rg_eqs}. 


\section{Mean-field phase diagram in the presence of SOC}\label{sec:mean_field}

The SOC contribution to the free energy given in Eq.~(\ref{eq:soc_free_energy}) also plays an important role in determining the type of magnetic order that develops at the magnetic phase transition. Consider $T\rightarrow T_{\rm mag}$: for $\alpha_1 < \alpha_2,\alpha_3$ (or $\alpha_2 < \alpha_1,\alpha_3$) only the SSDW and SVC phases can occur, while if $\alpha_3 < \alpha_1,\alpha_2$ only the SSDW and CSDW phases are possible. In the presence of SOC the mean-field magnetic transition temperature is shifted, leading to
\begin{eqnarray}
	\widetilde{T}_{{\rm mag}} = T_{\rm mag} - \frac{\text{min}\{\alpha_i\}}{a}\,.
\end{eqnarray}
Hence, if $\alpha_1 < \alpha_2,\alpha_3$, only $M_{x,1}$ and $M_{y,2}$ can condense in the vicinity of $\widetilde{T}_{{\rm mag}}$ thus leading to either an SSDW or SVC phase. This is at the core of the frustration mentioned above. The ground state in the spin \emph{isotropic} case can be incompatible with the moment direction enforced by the SOC. For instance, in the case above, the CSDW phase is ruled out by the SOC (see also Fig.~\ref{fig:case_a}). Note that only the case where the spin isotropic ground state is $C_4$ symmetric can lead to frustration. The $C_2$ symmetric SSDW phase is allowed regardless of whether the anisotropy is in-plane or out-of-plane. As $T \rightarrow \widetilde{T}_{\rm mag}$ the quadratic coefficients of the action will decide the type of magnetic order by imposing a certain direction of the magnetic moments. At lower temperatures, $T < \widetilde{T}_{\rm mag}$, the quartic coefficients become important. In cases where these are incompatible with the magnetic moment direction imposed by the SOC, additional phases can appear in an effort to lift the resulting frustration. These additional phases are mixtures of the three well-known phases, SSDW, CSDW, and SVC.

\subsection{Strong spin anisotropy}

\begin{figure}
\centering
\includegraphics[width=0.75\columnwidth]{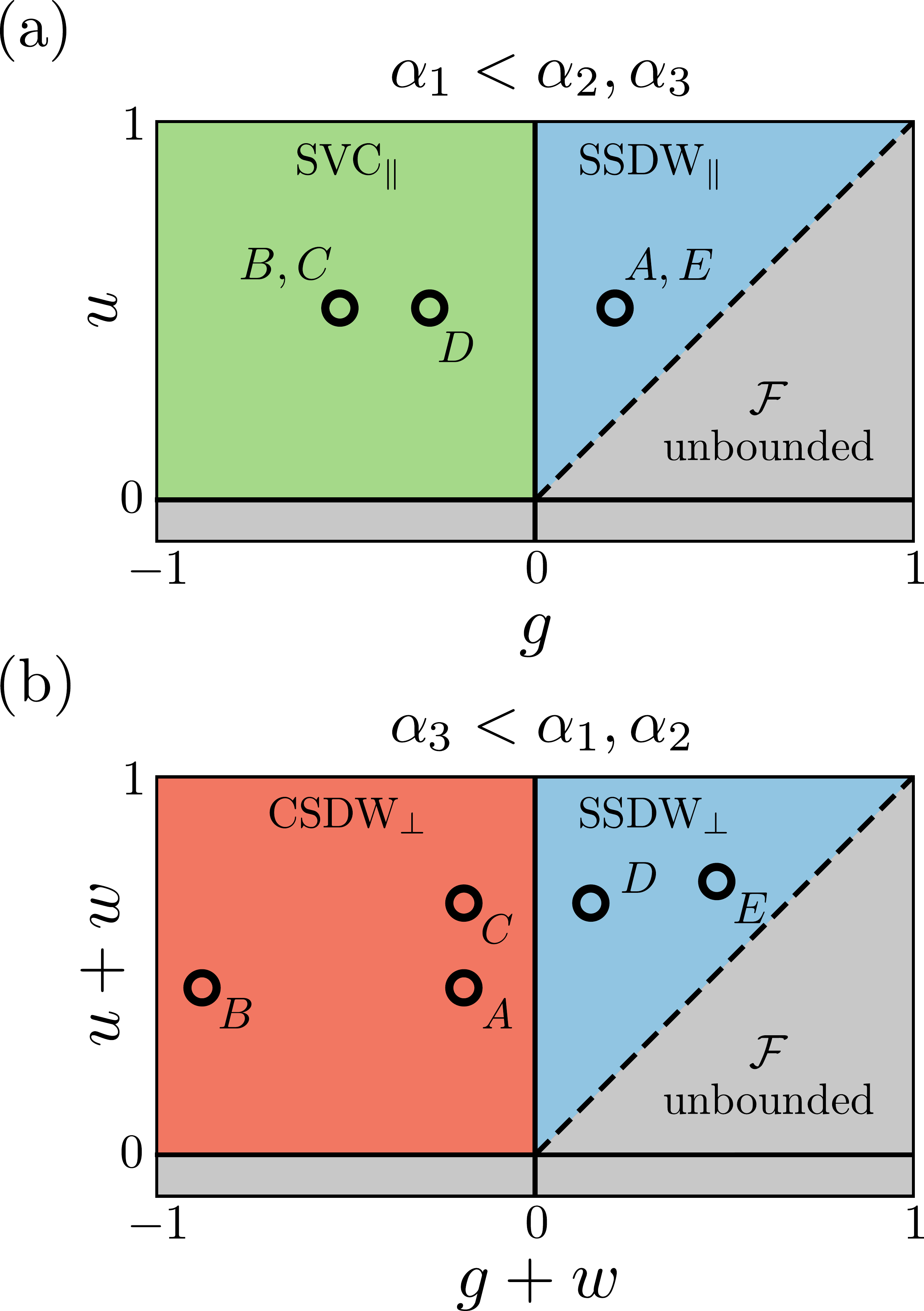}
\caption{\label{fig:anisotropic_phase_diagrams}(Color online) Mean-field phases immediately below $\widetilde{T}_{\rm mag}$ when (a) $\alpha_1 < \alpha_2,\alpha_3$ and (b) $\alpha_3 < \alpha_1,\alpha_2$. The points $\{A,B,C,D,E\}$ correspond to the ones depicted in Fig.~\ref{fig:mf_phase_diagram_w_points}. Gray areas are regions where the free energy is unbounded, which are not covered within our current approximation. Here $\perp$ refers to a out-of-plane moment direction (along $z$), while $\parallel$ refers to an in-plane moment direction.}
\end{figure}

Before discussing the general case of the full free energy in the presence of SOC let us first consider the limiting cases of $\alpha_1\ll\alpha_2,\alpha_3$ and $\alpha_3\ll\alpha_1,\alpha_2$. The case $\alpha_2\ll\alpha_1,\alpha_3$ is analogous to the first one and the same phases appear, only with moments pointing in different directions. In the case of $\alpha_1\ll\alpha_2,\alpha_3$ the degrees of freedom $M_{y,1}$, $M_{x,2}$, $M_{z,1}$, and $M_{z,2}$ are quenched and the free energy takes the simple form
\begin{eqnarray}
	\mathcal{F}_{\alpha_1} &=& \frac{1}{2}\int_q \left(r_0 + \alpha_1 \right)\left( M_{x,1}^2 + M_{y,2}^2 \right) \nonumber \\ &+& \frac{u}{2}\int_{x'} \left( M_{x,1}^2 + M_{y,2}^2 \right)^2 \nonumber \\ &-& \frac{g}{2}\int_{x'} \left( M_{x,1}^2 - M_{y,2}^2 \right)^2\,, \label{eq:free_energy_alpha1}
\end{eqnarray}
Note that the $w$-term drops out as the only non-zero spin components are $M_{x,1}$ and $M_{y,2}$. Evidently, when $g > 0$ an SSDW$_{\parallel}$ phase emerges, while if $g < 0$ an SVC$_{\parallel}$ phase is preferred. Here and throughout we use $\parallel$ to refer to  phases with in-plane moments and $\perp$ to phases with out-of-plane moments. Additionally, to ensure a bounded free energy, we require $u>g$ in the former case, while $u>0$ in the latter. This is summarized in Fig.~\ref{fig:anisotropic_phase_diagrams}(a). We note that the CSDW phase is absent in this case.
Similarly, assuming $\alpha_3\ll\alpha_1,\alpha_2$, the spin components along both $x$- and $y$-directions, $M_{x,1}$, $M_{x,2}$, $M_{y,1}$, and $M_{y,2}$, are quenched, and the free energy can be written as
\begin{eqnarray}
	\mathcal{F}_{\alpha_3} &=& \frac{1}{2}\int_k \left(r_0 + \alpha_3 \right)\left( M_{z,1}^2 + M_{z,2}^2 \right) \nonumber \\ &+& \frac{u+w}{2}\int_{x'} \left( M_{z,1}^2 + M_{z,2}^2 \right)^2 \nonumber \\ &-& \frac{g+w}{2}\int_{x'} \left( M_{z,1}^2 - M_{z,2}^2 \right)^2\,. \label{eq:free_energy_alpha3}
\end{eqnarray}
Here, an SSDW$_{\perp}$ phase appears for $g + w>0$, while boundedness again requires $u > g$. On the other hand, if $g + w < 0$ a $C_4$ phase emerges, although in this case it is a CSDW$_{\perp}$ phase. For $g + w < 0$ we require that $u + w>0$ for the free energy to remain bounded. Fig.~\ref{fig:anisotropic_phase_diagrams}(b) summarizes these findings. Here, the SVC phase is absent. The absence of one of the $C_4$-phases in the strongly anisotropic phase diagram is at the heart of the aforementioned frustration, i.e. the situation that quadratic and quartic coefficients favor two different types of magnetic order. In the presence of SOC and in the immediate vicinity of $\widetilde{T}_{{\rm mag}}$, the leading instabilities are not determined by the spin isotropic phase diagram of Fig.~\ref{fig:mf_phase_diagram_w_points} but rather by the spin anisotropic phase diagrams of Fig.~\ref{fig:anisotropic_phase_diagrams}.

\subsection{Moderate to weak spin anisotropy}

We proceed to consider temperatures well below $\widetilde{T}_{\rm mag}$ and in this way move beyond the leading instabilities. We examine several values of the quartic coefficients, indicated by the points $\{A,B,C,D,E\}$ shown in Figs.~\ref{fig:mf_phase_diagram_w_points} and \ref{fig:anisotropic_phase_diagrams}. With these we can construct simple mean-field phase diagrams as functions of temperature and the ratios of $\alpha_1$, $\alpha_2$ and $\alpha_3$. As discussed above, the case $\alpha_2 \ll \alpha_1,\alpha_3$ is analogous to $\alpha_1\ll\alpha_2,\alpha_3$ and will not be mentioned separately below.

A full analytical minimization of the free energy is difficult in the absence of spin rotational invariance. Instead, we carry out a numerical minimization of the free energy in the cases $\{A,B,C,D,E\}$. In case $A$ we supplement the discussion of the numerical results by an analytical treatment based on physically motivated expressions for the magnetic order parameters. While such a study is possible in all five cases, we focus on one of them for brevity, as this is sufficient to convey the main idea.

Throughout this section we vary the ratio of $\alpha_1/\alpha_3$ with fixed $\alpha_2$. In practice this is accomplished by fixing $\alpha_3/u=0.03$ and varying $\alpha_1$. Below we consider two cases. In one case, $\alpha_1,\alpha_3\ll\alpha_2$, and we take the limit $\alpha_2\rightarrow \infty$ (moderate anisotropy). The spin components associated with $\alpha_2$ can thus be safely ignored. In the other case, $\alpha_1,\alpha_3\lesssim\alpha_2$, and we take $\alpha_2/u=0.06$ (weak anisotropy). In this situation the spin components associated with $\alpha_2$ will affect the magnetic phase diagram, as we illustrate below. We take $a/u=5$.

\subsubsection{Parameter set $A$}

\begin{figure*}
\centering
\includegraphics[width=\textwidth]{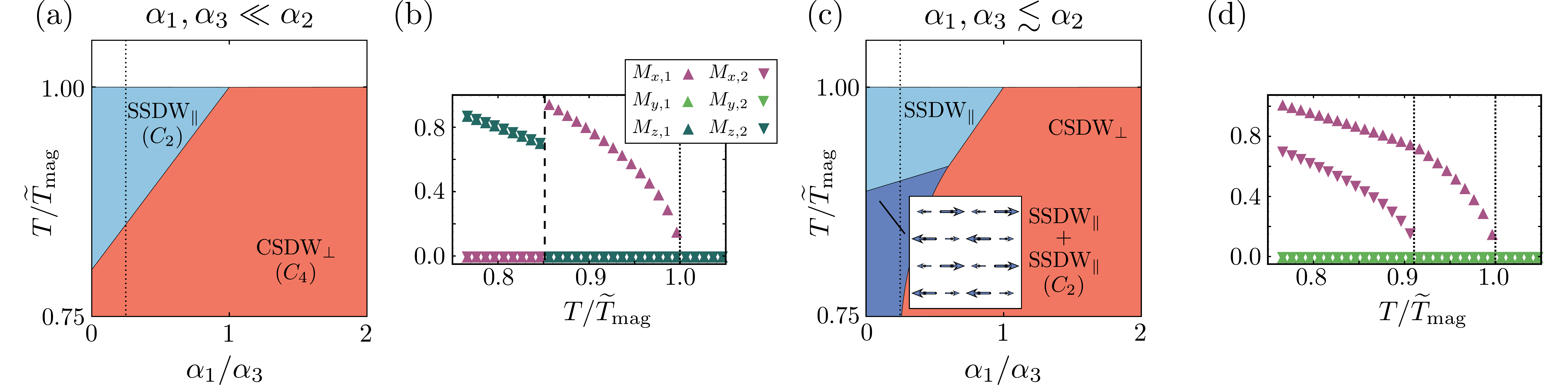}
\caption[]{\label{fig:case_a}(Color online) Mean-field phase diagrams for $g/u=0.20$ and $w/u=-0.25$, corresponding to point $A$ in Fig.~\ref{fig:mf_phase_diagram_w_points} for (a) $\alpha_1,\alpha_3\ll\alpha_2$ and (c) $\alpha_1,\alpha_3\lesssim\alpha_2$ respectively. The colors \tikz\draw[ssdw,fill=ssdw] (0,0) circle (.5ex); and \tikz\draw[csdw,fill=csdw] (0,0) circle (.5ex); refer to the SSDW and CSDW phases, while \tikz\draw[ssdwpcsdw,fill=ssdwpcsdw] (0,0) circle (.5ex); refer to the mixed phase SSDW$_{\parallel}$+SSDW$_{\parallel}$. In (b) and (d) we depict the temperature evolution of the order parameters for a constant $\alpha_1/\alpha_3=0.25$ indicated by the dotted lines in (a) and (c). Phase transitions are denoted by lines in (b) and (d), dotted lines are second-order while dashed lines are first-order. The in-plane to out-of-plane transition is seen to be first-order.}
\end{figure*}
As depicted in Fig.~\ref{fig:mf_phase_diagram_w_points}, the parameter set $A$ corresponds to $g/u=0.20$ and $w/u=-0.25$, which would predict a CSDW phase in the spin isotropic case. Indeed, if $\alpha_3>\alpha_1,\alpha_2$ this agrees with the numerical results, presented in Fig.~\ref{fig:case_a}.
On the other hand, if $\alpha_1<\alpha_2,\alpha_3$ an SSDW phase is found. This can be understood from Fig.~\ref{fig:anisotropic_phase_diagrams}. When $\alpha_3<\alpha_1,\alpha_2$, point $A$ is found in the CSDW phase, while for $\alpha_1<\alpha_2,\alpha_3$, it lies in the SSDW phase. At lower temperatures however, the behavior is vastly different, and depends on the size of $\alpha_2$. If the spin components associated with $\alpha_2$ can be ignored (i.e. $\alpha_1,\alpha_3 \ll \alpha_2$) the in-plane SSDW phase undergoes a first-order transition to an out-of-plane CSDW phase [Fig.~\ref{fig:case_a}(c)]. On the other hand, for $\alpha_1,\alpha_3\lesssim\alpha_2$, the spin components associated with $\alpha_2$ can condense. In this case an additional in-plane SSDW phase appears, albeit with moments aligned perpendicular to the ordering vector. The resulting phase is thus a superposition of two in-plane SSDW phases with $|\mbf{M}_1| \neq |\mbf{M}_2|$, and is depicted in the inset of Fig.~\ref{fig:case_a}(b). The transition from a single SSDW phase to a superposition of two is second-order, as seen in Fig.~\ref{fig:case_a}(d).

To understand these observations in further detail we consider the free energy for a number of different types of magnetic order. We begin with the case in which $\alpha_1,\alpha_3\ll\alpha_2$, in which case the spin components associated with $\alpha_2$ are quenched. The leading instabilities associated with Figs.~\ref{fig:anisotropic_phase_diagrams}(a) and (b) naturally provide two such magnetic orders. These are respectively the in-plane SSDW$_{\parallel}$ and the out-of-plane CSDW$_{\perp}$ phase. As temperature is lowered we must entertain the possibility that these two phases mix, yielding SSDW$_{\parallel}+$CSDW$_{\perp}$. The quartic coefficients preclude the appearance of an SVC phase, as evidenced in Figs.~\ref{fig:anisotropic_phase_diagrams}(a) and (b). Additionally, an in-plane CSDW$_{\parallel}$ phase is forbidden as the spin components associated with $\alpha_2$, i.e. $M_{y,1}$ and $M_{x,2}$, are quenched. Likewise, an out-of-plane SSDW$_{\perp}$ phase will always have a higher free energy compared to an out-of-plane CSDW$_{\perp}$ phase due to the quartic coefficients. We are thus left with just three expressions for the magnetic order parameters:
\begin{eqnarray}
	\text{SSDW$_{\parallel}$:} \quad && \begin{cases} \mbf{M}_1 = (M_{\rm SSDW},0,0) \\ \mbf{M}_2 = (0,0,0)\end{cases} \\
	\text{CSDW$_{\perp}$:} \quad && \begin{cases} \mbf{M}_1 = (0,0,M_{\rm CSDW}) \\ \mbf{M}_2 = (0,0,M_{\rm CSDW}) \end{cases} \\
	\text{SSDW$_{\parallel}$+CSDW$_{\perp}$:} \quad && \begin{cases} \mbf{M}_1 = (M_{\rm SSDW},0,M_{\rm CSDW}) \\ \mbf{M}_2 = (0,0,M_{\rm CSDW}) \end{cases}\,.
\end{eqnarray}
Let us first consider the mixed case. The free energy is
\begin{eqnarray}
	\mathcal{F}_{\rm SSDW_{\parallel} + CSDW_{\perp}} &=& \frac{1}{2}(r_0 + \alpha_1)M_{\rm SSDW}^2 + (r_0 + \alpha_3)M_{\rm CSDW}^2 \nonumber \\
	&+& \frac{1}{2}(u-g)M_{\rm SSDW}^4 + 2(u+w) M_{\rm CSDW}^4 \nonumber \\ &+& 2 u M^2_{\rm SSDW} M^2_{\rm CSDW}\,.\label{eq:mixed_ssdw_par_csdw_perp}
\end{eqnarray}
A coupling between $M_{\rm SSDW}$ and $M_{\rm CSDW}$ arises from the term $\tfrac{u}{2}(\mbf{M}_1^2 + \mbf{M}_2^2)^2$, while the remaining terms only involve either $M_{\rm SSDW}$ or $M_{\rm CSDW}$. Due to the restrictions imposed on $u$ to ensure a bounded free energy [cf. the discussion following Eq.~(\ref{eq:part_func})], the energy cost of the crossterm will always outweigh the energy gained from the remaining coefficients. This can be understood from a general comparison of the quartic coefficients. For Eq.~(\ref{eq:mixed_ssdw_par_csdw_perp}), a coexistence phase is possibly only if
\begin{eqnarray}
	u(w-g)>gw\,,
\end{eqnarray}
which cannot be satisfied for any $u$ simultaneously fulfilling $u>g$ and $u>-w$. Thus, when $\alpha_1,\alpha_3 \ll \alpha_2$ no mixed phase will exist, regardless of the temperature or the value of $\alpha_1/\alpha_3$. For the SSDW and CSDW cases we find
\begin{eqnarray}
	M_{\rm SSDW} &=& \frac{1}{\sqrt{2}}\sqrt{-\frac{r_0 + \alpha_1}{u-g}}\,,\\
	M_{\rm CSDW} &=& \frac{1}{2}\sqrt{-\frac{r_0 + \alpha_3}{u+w}}\,,
\end{eqnarray}
and the final expressions for the free energies are
\begin{eqnarray}
	\mathcal{F}_{\rm SSDW_{\parallel}} &=& -\frac{1}{8}\frac{(r_0+\alpha_1)^2}{u-g}\,, \\
	\mathcal{F}_{\rm CSDW_{\perp}} &=& -\frac{1}{8}\frac{(r_0+\alpha_3)^2}{u+w}\,.
\end{eqnarray}
Recall $r_0 = a(T-T_{\rm mag})$ and we can compare the above expressions along with the free energy for the paramagnetic state, $\mathcal{F}_{\rm PM}=0$. This yields lines of primary and secondary transitions identical to those obtained from the numerical results presented in Fig.~\ref{fig:case_a}(a).

Let us now analytically consider the case $\alpha_1,\alpha_3\lesssim\alpha_2$, implying that the spin components associated with $\alpha_2$ cannot be set to zero. This situation is slightly more complicated due to the presence of these additional spin components. The mixed phase SSDW$_{\parallel}+$CSDW$_{\perp}$ is ruled out by arguments identical to the ones presented above, i.e. the energy cost of the crossterm outweighs the energy gained from the remaining coefficients. However, when $\alpha_2$ is comparable to $\alpha_1$ and $\alpha_3$, a phase consisting of the superposition of an SSDW$_{\parallel}$ phase with moments parallel to the ordering vector and an SSDW$_{\parallel}$ phase with moments perpendicular to the ordering vector must be considered. In addition, the in-plane SSDW$_{\parallel}$ phase and the out-of-plane CSDW$_{\perp}$ phase are expected to be present. Once again, any SVC phases are precluded due to the choice of quartic coefficients. Hence we start from the expressions
\begin{eqnarray}
	\text{SSDW$_{\parallel}$:} \quad && \begin{cases} \mbf{M}_1 = (M_{\rm SSDW},0,0) \\ \mbf{M}_2 = (0,0,0)\end{cases} \\
	\text{CSDW$_{\perp}$:} \quad && \begin{cases} \mbf{M}_1 = (0,0,M_{\rm CSDW}) \\ \mbf{M}_2 = (0,0,M_{\rm CSDW}) \end{cases} \\
	\text{SSDW$_{\parallel}$+SSDW$_{\parallel}$:} \quad && \begin{cases} \mbf{M}_1 = (M_{\rm SSDW_1},0,0) \\ \mbf{M}_2 = (M_{\rm SSDW_2},0,0) \end{cases}\,. \label{eq:ssdw+ssdw_ansatz}
\end{eqnarray}
In constrast to the case above, the free energy for the mixed phase contains multiple terms coupling $M_{\rm SSDW_1}$ and $M_{\rm SSDW_2}$:
\begin{eqnarray}
	&&\mathcal{F}_{\rm SSDW_{\parallel}+SSDW_{\parallel}} = \nonumber \\ && \qquad \frac{1}{2}(r_0 + \alpha_1)M_{\rm SSDW_1}^2 + \frac{1}{2}(r_0 + \alpha_2) M_{\rm SSDW_2}^2 \nonumber \\
	&& \qquad + \frac{1}{2}(u-g)M^{4}_{\rm SSDW_1} + \frac{1}{2}(u-g)M^{4}_{\rm SSDW_2} \nonumber \\ && \qquad + (u + g + 2w)M^{2}_{\rm SSDW_1} M^{2}_{\rm SSDW_2}
\end{eqnarray}
In contrast to the case with SSDW$_{\parallel}$+CSDW$_{\perp}$ the system can now take advantage of the cross term due to the presence of the additional coefficients $g$ and $w$. In this case, a coexistence phase is possible if
\begin{eqnarray}
	g+w<0\,.
\end{eqnarray}
Evidently, this is satisfied in the triangle in which $A$ is located in Fig.~\ref{fig:mf_phase_diagram_w_points}. Hence an SSDW$_{\parallel}$+SSDW$_{\parallel}$ phase is energetically favorable in a region of parameter space. A comparison of the free energies of the three phases confirms the results of the numerical minimization, presented in Fig.~\ref{fig:case_a}(b). We note that there is some ambiguity associated with the naming of the phase SSDW$_{\parallel}$+SSDW$_{\parallel}$; we could equally well have denoted it by SSDW$_{\parallel}$+CSDW$_{\parallel}$ as is also clear from Eq.~(\ref{eq:ssdw+ssdw_ansatz}) and Fig.~\ref{fig:case_a}(d).

\subsubsection{Parameter set $B$}

\begin{figure*}
\centering
\includegraphics[width=\textwidth]{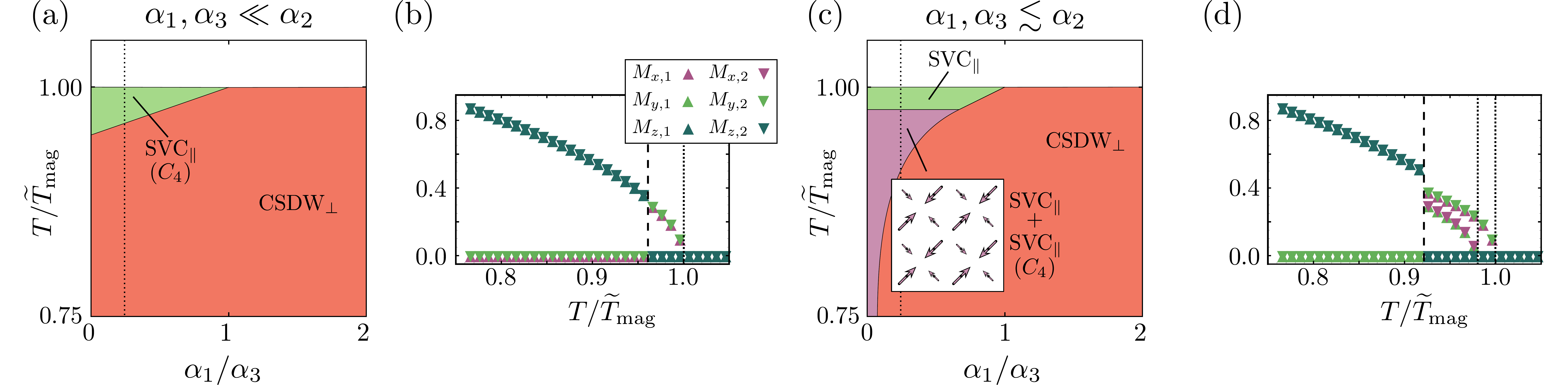}
\caption[]{\label{fig:case_b}(Color online) Mean-field phase diagrams for $g/u=-0.25$ and $w/u=-0.25$ corresponding to point $B$ in Fig.~\ref{fig:mf_phase_diagram_w_points} for (a) $\alpha_1,\alpha_3\ll\alpha_2$ and (c) $\alpha_1,\alpha_3\lesssim\alpha_2$ respectively. The colors \tikz\draw[svc,fill=svc] (0,0) circle (.5ex); and \tikz\draw[csdw,fill=csdw] (0,0) circle (.5ex); refer to the SVC and CSDW phases, while \tikz\draw[csdwpsvc,fill=csdwpsvc] (0,0) circle (.5ex); refers to the SVC$_{\parallel}$+SVC$_{\parallel}$ phase appearing when the spin components associated with $\alpha_2$ can condense. This phase is depicted in the inset of (c) and is a superposition of the hedgehog- and loop-SVC phases. The temperature evolution of the order parameters for a specific choice of $\alpha_1/\alpha_3=0.25$ [corresponding to the dotted in line in (a) and (c)] is shown in (b) and (d). Phase transitions are denoted by lines in (b) and (d), dotted lines are second-order transitions, while dashed lines are first-order.}
\end{figure*}

For this point, the quartic coefficients are $g/u=-0.25$ and $w/u=-0.25$. Similarly to case $A$ this predicts a CSDW phase in the spin isotropic case. The difference comes from the location of the point $B$ in Fig.~\ref{fig:anisotropic_phase_diagrams}. For $\alpha_1<\alpha_2,\alpha_3$ we find point $B$ in the SVC$_{\parallel}$ phase in contrast to the point $A$, which was in the SSDW$_{\parallel}$ phase. On the other hand, for $\alpha_3<\alpha_1,\alpha_2$, $B$ remains in the CSDW$_{\perp}$ phase. The result of numerically minimizing the free energy in this case is depicted in Fig.~\ref{fig:case_b} for both $\alpha_1,\alpha_3\ll\alpha_2$ and $\alpha_1,\alpha_3\lesssim\alpha_2$.
These results can be understood from arguments similar to the ones presented for case $A$ above. From the phase diagrams in Fig.~\ref{fig:anisotropic_phase_diagrams} we expect to find both an SVC$_{\parallel}$ phase and a CSDW$_{\perp}$ phase, which are indeed found in Figs.~\ref{fig:case_b}(a) and (b). For the case $\alpha_1,\alpha_3\lesssim\alpha_2$ a third phase is uncovered, consisting of two intertwined hedgehog- and loop-SVC$_{\parallel}$ phases~\cite{meier17,halloran17}. The magnetization profile of this phase is depicted in the inset in Fig.~\ref{fig:case_b}(b). One might expect a CSDW$_{\parallel}$+SVC$_{\parallel}$ phase to occur in addition to, or in place of, the SVC$_{\parallel}$+SVC$_{\parallel}$ phase. In fact, a CSDW$_{\parallel}$+SVC$_{\parallel}$ phase would break the tetragonal symmetry. This incurs a penalty since $g<0$ makes such a phase unfavorable compared to the SVC$_{\parallel}$+SVC$_{\parallel}$.

Figs.~\ref{fig:case_b}(c) and (d) depict the evolution of the order parameters for $\alpha_1/\alpha_3=0.25$ for both $\alpha_1,\alpha_3\ll\alpha_2$ and $\alpha_1,\alpha_3\lesssim\alpha_2$. The first-order transitions between in-plane and out-of-plane phases observed for point $A$ above are also evident in the cases presented here.

\subsubsection{Parameter set $C$}

\begin{figure*}
\centering
\includegraphics[width=1\textwidth]{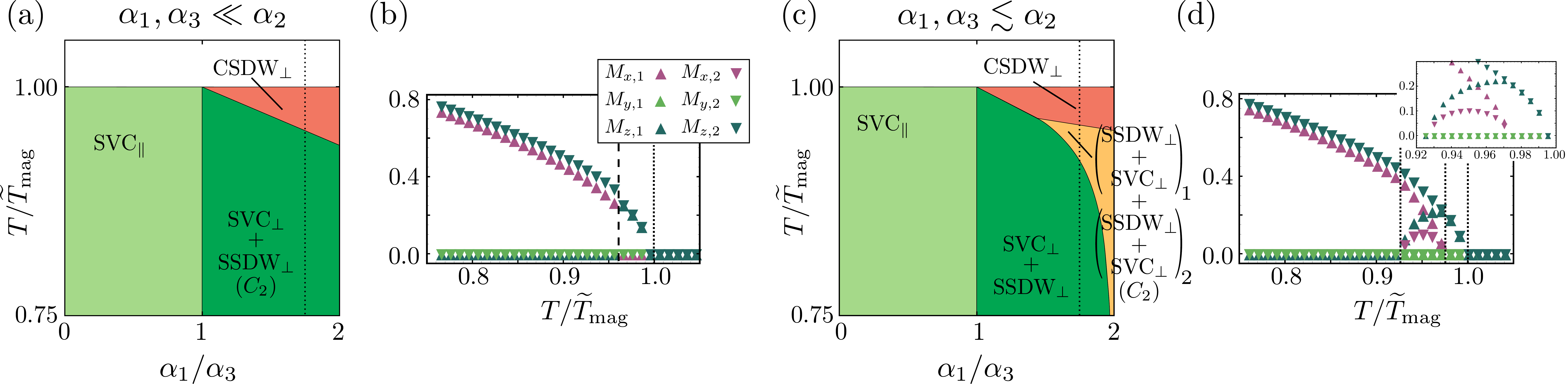}
\caption[]{\label{fig:case_c}(Color online) Mean-field phase diagrams for $g/u=-0.25$ and $w/u=0.125$ corresponding to point $C$ in Fig.~\ref{fig:mf_phase_diagram_w_points} for (a) $\alpha_1,\alpha_3\ll\alpha_2$ and (c) $\alpha_1,\alpha_3\lesssim\alpha_2$. Dark green (\tikz\draw[ssdwpsvc,fill=ssdwpsvc] (0,0) circle (.5ex);) refers to an SVC$_{\perp}$+SSDW$_{\perp}$ phase dominating for $\alpha_1/\alpha_3>1$. The yellow (\tikz\draw[ssdwpcsdwpsvc,fill=ssdwpcsdwpsvc] (0,0) circle (.5ex);) area denotes a region hosting two intertwined SVC$_{\perp}$+SSDW$_{\perp}$ phases, which only appears for $\alpha_1,\alpha_3\lesssim\alpha_2$. The temperature evolution of the order parameters is depicted in (b) and (d) for $\alpha_1/\alpha_3=1.75$ corresponding to the dotted lines in (a) and (c). The transition between CSDW$_{\perp}$ and SVC$_{\perp}$+SSDW$_{\perp}$ features a reorientation of the magnetic moments and is first-order. In contrast, if $\alpha_1,\alpha_3\lesssim\alpha_2$ the system can go from CSDW$_{\perp}$ to SVC$_{\perp}$+SSDW$_{\perp}$ through an intermediary phase, which is accessed through second-order transitions, see the inset in (d). Phase transitions are denoted by lines in (b) and (d), dotted lines indicate second-order transitions while dashed lines are first-order transitions.}
\end{figure*}

The quartic coefficients in this case are $g/u=-0.25$ and $w/u=0.125$. In the spin isotropic case they give an SVC phase. In the presence of spin anisotropy with $\alpha_1<\alpha_2,\alpha_3$ this matches expectations based on Fig.~\ref{fig:anisotropic_phase_diagrams}(a). This is in contrast to the case $\alpha_3<\alpha_1,\alpha_2$ in which a CSDW$_{\perp}$ phase is expected. These two phases indeed appear as leading instabilities, as seen from the numerical phase diagram presented in Fig.~\ref{fig:case_c}.
As temperature is lowered an SVC$_{\perp}$+SSDW$_{\perp}$ phase appears, breaking the tetragonal symmetry. Such a phase appears as the system attempts to accommodate an SVC phase. For $\alpha_1/\alpha_3>1$ the preferred phase has out-of-plane components, initially leading to the CSDW$_{\perp}$ phase. With lower temperatures the system can gain energy by developing components along both in-plane and out-of-plane directions:
\begin{eqnarray}
	\text{SVC$_{\perp}$+SSDW$_{\perp}$:} \quad && \begin{cases} \mbf{M}_1 = (M_{\rm SVC},0,0) \\ \mbf{M}_2 = (0,0,M_{\rm SVC} + M_{\rm SSDW})\end{cases}\,.
\end{eqnarray}
Here we have chosen the state in which $M_{x,1}$ condenses. The state with non-zero $M_{y,2}$ (and $M_{z,1}$) is related to the above by a $C_4$ rotation. The system spontaneously selects one of the two. The SVC components are favored by the quartic coefficients while the SSDW component appears as the $x$- and $z$-components cannot be identical due to the spin anisotropy. Evidently, this phase can be seen as a modified out-of-plane SVC phase with $|\mbf{M}_1|\neq|\mbf{M}_2|$. Note that out-of-plane SVC phases with $|\mbf{M}_1|=|\mbf{M}_2|$ can only exist for fine-tuned spin anisotropies $\alpha_3=\alpha_1$ (or $\alpha_3=\alpha_2$).

These arguments also apply to the case where $\alpha_1,\alpha_3\lesssim\alpha_2$. In this case however, the system exploits the presence of an additional soft in-plane direction to form two intertwined out-of-plane SVC$_{\perp}$+SSDW$_{\perp}$ phases. One is the same as appears when $\alpha_1,\alpha_3\ll\alpha_2$. The second SVC$_{\perp}$+SSDW$_{\perp}$ phase has the form
\begin{eqnarray}
	\text{SVC$_{\perp}$+SSDW$_{\perp}$:} \quad && \begin{cases} \mbf{M}_1 = (0,0,M_{\rm SVC}+ M_{\rm SSDW}) \\ \mbf{M}_2 = (M_{\rm SVC},0,0 )\end{cases}\,,
\end{eqnarray}
and arises as the system attempts to balance the finite $\alpha_2$, which allows for non-zero $M_{x,2}$ and $M_{y,1}$, with the fact that $w<|g|$. The relative size of the quartic coefficients implies that the system prioritizes minimizing $\mbf{M}_1^2 - \mbf{M}_2^2$ over $\mbf{M}_1 \cdot \mbf{M}_2$. For spin anisotropic systems a balance is struck between the quadratic and quartic coefficients. The result is that the system can gain energy by having both $\mbf{M}_1^2 - \mbf{M}_2^2 \neq 0$ and $\mbf{M}_1\cdot\mbf{M}_2\neq0$. We therefore find the total magnetic order parameter in the yellow (\tikz\draw[ssdwpcsdwpsvc,fill=ssdwpcsdwpsvc] (0,0) circle (.5ex);) region in Fig.~\ref{fig:case_c}(b) to be
\begin{eqnarray}
	&&\text{(SVC$_{\perp}$+SSDW$_{\perp}$)$_{1}$+(SVC$_{\perp}$+SSDW$_{\perp}$)$_{2}$:} \nonumber \\ && \qquad
	\begin{cases} \mbf{M}_1 = (M_{\rm SVC_1},0,M_{\rm SVC_2}+ M_{\rm SSDW_2}) \\ \mbf{M}_2 = (M_{\rm SVC_2},0,M_{\rm SVC_1}+M_{\rm SSDW_1})\end{cases}\,.
\end{eqnarray}
The intricate evolution of this order as a function of temperature is captured in the inset of Fig.~\ref{fig:case_c}(d). In this case the moments do not reorient as they did in the cases $A$ and $B$ above. Instead, the in-plane components condense via a second-order phase transition while the out-of-plane components split, with one going smoothly to zero. This is in contrast to the direct transition between the CSDW$_{\perp}$ and SVC$_{\perp}$+SSDW$_{\perp}$ phases depicted in Fig.~\ref{fig:case_c}(c). In this case, the in-plane component condenses through a first-order transition. Simultaneously, one of the out-of-plane components drops to zero.

\subsubsection{Parameter set $D$}

\begin{figure*}
\centering
\includegraphics[width=\textwidth]{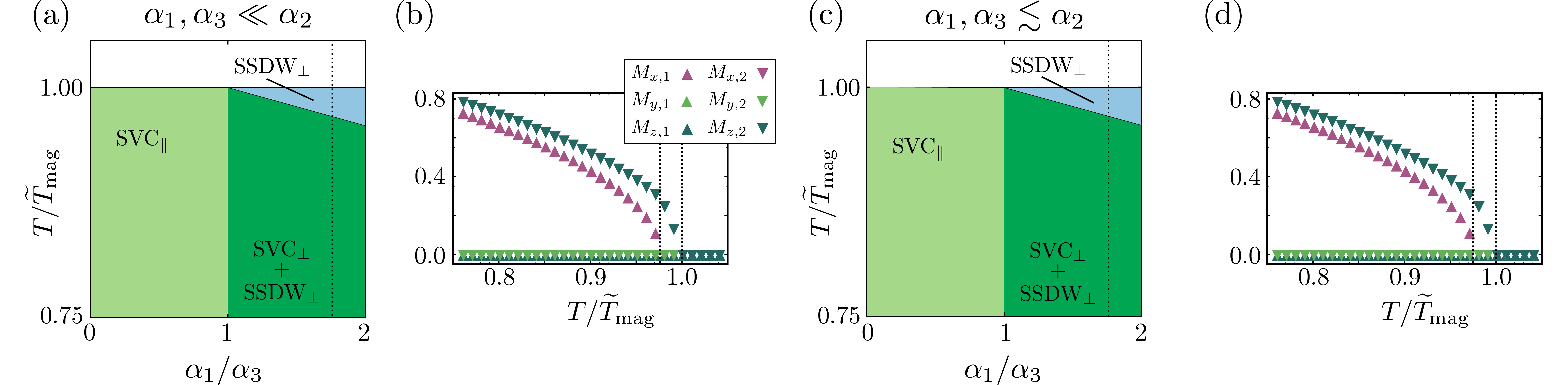}
\caption{\label{fig:case_d}(Color online) Mean-field phase diagrams for $g/u=-0.125$ and $w/u=0.25$ corresponding to point $D$ in Fig.~\ref{fig:mf_phase_diagram_w_points} for (a) $\alpha_1,\alpha_3\ll\alpha_2$ and (c) $\alpha_1,\alpha_3\lesssim\alpha_2$. In (b) and (d) we show the order parameters as a function of temperature for $\alpha_1/\alpha_3=1.75$ corresponding to the dotted lines in (a) and (c). We find a second-order transition between the SSDW$_{\perp}$ phase and the SVC$_{\perp}$+SSDW$_{\perp}$ phase. In (b) and (d) dotted lines denote second-order transitions between phases.}
\end{figure*}

Here the quartic coefficients are $g/u=-0.125$ and $w/u=0.25$, also indicating an SVC phase in the spin isotropic case. However, the relative magnitude of $g$ and $w$ play an important role, as is evident from Fig.~\ref{fig:anisotropic_phase_diagrams}. As in the case studied above, for $\alpha_1<\alpha_2,\alpha_3$ point $D$ is in the SVC$_{\parallel}$ phase. However, for $\alpha_3<\alpha_1,\alpha_2$ point $D$ is found in the SSDW$_{\perp}$ phase. This fact is reflected in the numerical phase diagrams presented in Fig.~\ref{fig:case_d}.
The appearance of the SVC$_{\perp}$+SSDW$_{\perp}$ phase in this case can be understood from arguments similar to those presented for case $C$ above. The main distinction to case $C$ is in the appearance of an SSDW$_{\perp}$ phase for $\alpha_1/\alpha_3>1$ at temperatures close to $\widetilde{T}_{\rm mag}$. As temperature is lowered this phase evolves to an SVC$_{\perp}$+SSDW$_{\perp}$ through a second-order phase transition, see Fig.~\ref{fig:case_d}(c) and (d). This occurs as the system attempts to reconcile the quadratic and the quartic coefficients. The quartic coefficients prefer an SVC phase, while the spin anisotropy prefers a dominant out-of-plane component. This prevents an SVC phase with $|\mbf{M}_1|=|\mbf{M}_2|$, and instead yields an SVC$_{\perp}$+SSDW$_{\perp}$ phase. In contrast to case $C$, the presence of $\alpha_2$ plays no role here and the two phase diagrams in Figs.~\ref{fig:case_d}(a) and (b) are identical. This is ultimately a consequence of the fact that $w>|g|$ which implies that the system can gain energy by remaining in a configuration for which $\mbf{M}_1 \cdot \mbf{M}_2 = 0$, but with $\mbf{M}_1^2 - \mbf{M}_2^2 \neq 0$.

\subsubsection{Parameter set $E$}

\begin{figure*}
\centering
\includegraphics[width=1\textwidth]{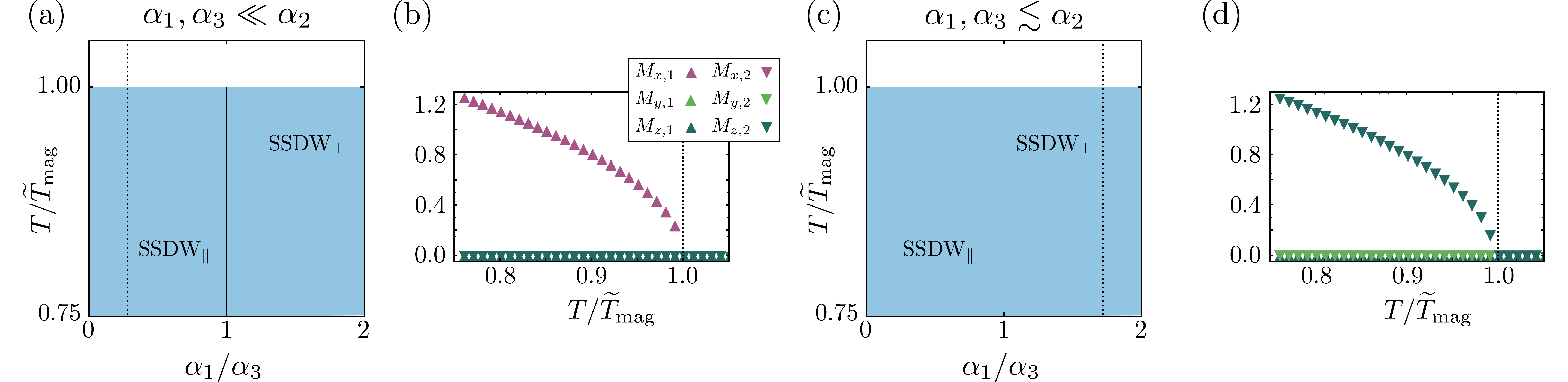}
\caption{\label{fig:case_e}(Color online) Mean-field phase diagrams for $g/u=0.25$ and $w/u=0.25$ corresponding to point $E$ in Fig.~\ref{fig:mf_phase_diagram_w_points} for (a) $\alpha_1,\alpha_3\ll\alpha_2$ and (c) $\alpha_1,\alpha_3\lesssim\alpha_2$. These are identical as all three types of spin anisotropy allow for an SSDW phase. In (b) and (d) the evolution of the order parameters as function of temperature is shown. In (b) $\alpha_1/\alpha_3=0.25$ and in (d) $\alpha_1/\alpha_3=1.75$. Hence, in (b) the moments are in-plane and (d) the moments are out-of-plane. The dotted lines in (b) and (d) denote second-order transitions.}
\end{figure*}

This point corresponds to $g/u=0.25$ and $w/u=0.25$ and for spin isotropic systems it lies deep in the SSDW phase. Generally, all points in the isotropic SSDW phase of Fig.~\ref{fig:mf_phase_diagram_w_points} map to the SSDW regions of Fig.~\ref{fig:anisotropic_phase_diagrams}. This is consistent with the fact that all types of spin anisotropies allow for the SSDW phase, and no frustration is anticipated in this case. This is confirmed by the numerical minimization of the free energy, as seen in Fig.~\ref{fig:case_e}. Unsurprisingly, the results for $\alpha_1,\alpha_3\ll\alpha_2$ and $\alpha_1,\alpha_3,\lesssim\alpha_2$ are identical.
As shown in Fig.~\ref{fig:case_e}(c) and (d) the order parameters also behave identically, although in (c) the moments are in-plane while in (d) they are out-of-plane.

\subsection{Summary of mean-field results}

The magnetic phase diagram is substantially modified in the presence of SOC, as revealed by Figs.~\ref{fig:case_a}--\ref{fig:case_e}. We find five regions of the isotropic mean-field phase diagram exhibiting distinct behavior. These regions are, respectively, (i) $g>0$ and $-w>g$, (ii) $g<0$ and $w<0$, (iii) $w>0$ and $-g>w$, (iv) $w>0$ and $-g<w$, and (v) $g>0$ and $-w<g$. Within each region we focus on a specific parameter set, leading to the five parameter sets $\{A,B,C,D,E\}$. As expected, SOC leads to a reorientation of the magnetic moments. Except for parameters that predict a SSDW in the isotropic case, it leads to frustration whose main consequences are twofold: First, as $T\rightarrow \widetilde{T}_{\rm mag}$ the phase is determined by the spin anisotropic phase diagrams in Fig.~\ref{fig:anisotropic_phase_diagrams}, rather than the spin isotropic one in Fig.~\ref{fig:mf_phase_diagram_w_points}. Second, at lower temperatures, the system seeks to balance the impact of the quadratic coefficients with the quartic ones. This leads to admixtures of the original three phases resulting in the appearance of a rich landscape of tetragonal and orthorhombic magnetic phases.

\section{Renormalization group analysis in presence of SOC}\label{sec:rg_eqs}

We now study the phase diagram beyond mean-field theory, employing an RG approach. The RG analysis is carried out at $T=0$ as the effects of spin anisotropy on the quartic terms are most pronounced there. In addition, at $T=0$ and $d=2$ the system lies at the upper critical dimension allowing for a well-controlled RG calculation. In Ref.~\onlinecite{christensen17b} we presented the main result of this treatment: the emergence of magnetic degeneracy for a wide range of initial bare parameters near the putative QCP. Here we provide further details for the derivation of the RG flow equations in the presence of spin anisotropy. Furthermore, we present the full numerical solutions of the RG flow equations, which are in agreement with the analytical treatment presented in Ref.~\onlinecite{christensen17b}.

For the purpose of deriving the RG equations, it is convenient to rewrite the action of Eq.~(\ref{eq:action_1}). Note that the anisotropy of the quadratic coefficients will generate anisotropic quartic coefficients under the RG flow, even if they are initially isotropic. The general form is
\begin{widetext}
\begin{eqnarray}
	\mathcal{S} &=& \frac{1}{2}\int_q\sum_{i}\left[ M_{i,1}(\mbf{q}) \left( \tilde{r}_{i,1} + q^2 \right) M_{i,1}(-\mbf{q}) + M_{i,2}(\mbf{q}) \left( \tilde{r}_{i,2} + q^2 \right) M_{i,2}(-\mbf{q})  \right] \nonumber \\ &+& \sum_{ij} \lambda^{ij}_1 \int_{q_1,q_2,q_3} M_{i,1}(\mbf{q}_1)M_{i,1}(\mbf{q}_2)M_{j,1}(\mbf{q}_3)M_{j,1}(-\mbf{q}_1-\mbf{q}_2-\mbf{q}_3) \nonumber \\
	&+& \sum_{ij} \lambda^{ij}_2 \int_{q_1,q_2,q_3} M_{i,2}(\mbf{q}_1)M_{i,2}(\mbf{q}_2)M_{j,2}(\mbf{q}_3)M_{j,2}(-\mbf{q}_1-\mbf{q}_2-\mbf{q}_3) \nonumber \\ 
	&+& 2\sum_{ij}\rho^{ij}\int_{q_1,q_2,q_3} M_{i,1}(\mbf{q}_1)M_{i,1}(\mbf{q}_2)M_{j,2}(\mbf{q}_3)M_{j,2}(-\mbf{q}_1-\mbf{q}_2-\mbf{q}_3) \nonumber \\
	&+& 2 \sum_{ij} w^{ij} \int_{q_1,q_2,q_3} M_{i,1}(\mbf{q}_1)M_{i,2}(\mbf{q}_2)M_{j,1}(\mbf{q}_3)M_{j,2}(-\mbf{q}_1-\mbf{q}_2-\mbf{q}_3)\,,\label{eq:ani_action}
\end{eqnarray}
\end{widetext}
here $i,j=x,y,z$ and we defined
\begin{eqnarray}
	\tilde{r}_{i,1} &=& r_0 + \delta_{ix}\alpha_1 + \delta_{iy}\alpha_2 + \delta_{iz}\alpha_3 \\
	\tilde{r}_{i,2} &=&  r_0 + \delta_{ix}\alpha_2 + \delta_{iy}\alpha_1 + \delta_{iz}\alpha_3\,.
\end{eqnarray} 
The quartic terms are written in momentum space. At $T=0$, the integrals $\int_q \equiv \int^{|\mbf{q}|<\Lambda}\frac{\mathrm{d}^4 q}{(2\pi)^4}$, as $d+z=2+2=4$. In this case, the $T=0$ Matsubara summation can be converted to a $2$-dimensional momentum integral, placing the system at the upper critical dimension. Here, $\Lambda$ is the upper cut-off. The indices of the quartic coefficients anticipates the fact that the anisotropy of the quadratic coefficients will generate anisotropies in the quartic coefficients. The coefficients $\lambda^{ij}_1$ and $\lambda_2^{ij}$ are related by $C_4$ symmetry:
\begin{eqnarray}
	\lambda_1^{ij} = \lambda_2^{\bar{i}\bar{j}}\,,
\end{eqnarray}
where $\bar{x}=y$, $\bar{y}=x$, and $\bar{z}=z$. Additionally, they are symmetric matrices
\begin{eqnarray}
	\lambda_{1,2}^{ij} = \lambda_{1,2}^{ji}\,.
\end{eqnarray}
In contrast, under a $C_4$ rotation $\rho^{ij}$ transforms according to
\begin{eqnarray}
	\rho^{ij} = \rho^{\bar{j}\bar{i}}\,,
\end{eqnarray}
i.e. $\rho^{xx}=\rho^{yy}$, but $\rho^{xy}$ and $\rho^{yx}$ are unrelated, see Eq.~(\ref{eq:ani_action}). $w^{ij}$ transforms according to
\begin{eqnarray}
	w^{ij} \rightarrow w^{\bar{i}\bar{j}}\,,
\end{eqnarray}
and is also symmetric:
\begin{eqnarray}
	w^{ij} = w^{ji}\,.
\end{eqnarray}
Hence, the number of independent quartic coefficients is 16, which, along with the three independent quadratic coefficients, yields a total of 19 coupled flow equations. Thus, despite the isotropic initial conditions imposed on the quartic coefficients,
\begin{eqnarray}
	\left(\lambda^{ij}_1 \right)_{(0)} &=& \left(\lambda^{ij}_2 \right)_{(0)}=\frac{u_{(0)}-g_{(0)}}{2}\,, \\ 
	\left(\rho^{ij}\right)_{(0)} &=& \frac{u_{(0)}+g_{(0)}}{2}\,,\\ 
	\left( w^{ij} \right)_{(0)} &=& w_{(0)}\,,
\end{eqnarray}
anisotropic terms are generated under the RG flow. Here we review the renormalization of the propagator and quartic vertices up to one-loop.

As a first step the magnetic degrees are separated into slow modes, $\mbf{M}^{<}$, and fast modes, $\mbf{M}^{>}$, i.e. $\mbf{M}(\mbf{q})=\mbf{M}^{<}(\mbf{q}) + \mbf{M}^{>}(\mbf{q})$ where
\begin{eqnarray}
	\mbf{M}^{<}(\mbf{q}) &=& 
	\begin{cases}
		\mbf{M}(\mbf{q}) & 0 \leq |\mbf{q}| \leq \Lambda e^{-\ell} \\
		0 & \text{otherwise}
	\end{cases}\,, \\
	\mbf{M}^{>}(\mbf{q}) &=& 
	\begin{cases}
		\mbf{M}(\mbf{q}) & \Lambda e^{-\ell} \leq |\mbf{q}| \leq \Lambda \\
		0 & \text{otherwise}
	\end{cases}\,,
\end{eqnarray}
and $\ell > 0$. The fast modes are integrated out yielding
\begin{eqnarray}
	Z &=& Z^{>}\int \mathcal{D}[M^{<}_{i,1},M^{<}_{i,2}]e^{-\mathcal{S}^{<}} \nonumber \\ && \qquad \times e^{-\left\langle \mathcal{S}_{\rm int} \right\rangle_{>,0} + \frac{1}{2}\left(\left\langle \mathcal{S}^{2}_{\rm int} \right\rangle_{>,0} - \left\langle \mathcal{S}_{\rm int} \right\rangle^2_{>,0} \right) + \cdots}\,,
\end{eqnarray}
where $\mathcal{S}_{\rm int}$ expresses how the high-momentum fast modes affect the relevant slow modes. Here $\langle \cdot \rangle_{>,0}$ refers to an average with respect to the Gaussian term of the fast modes. Terminating the expression at second order in $\mathcal{S}_{\rm int}$ corresponds to a one-loop approximation.

To ensure that the action describes the original physical system a subsequent momentum rescaling, $\mbf{q}=e^{\ell}\mbf{q}^{<}$, and field rescaling, $\mbf{M}^{<}(\mbf{q}^{<})=\zeta \mbf{M}(\mbf{q})$, is required. We follow the usual convention that the coefficient of the kinetic term $\mbf{q}^2$ should remain unchanged under such a rescaling. Considering the isotropic case for simplicity, we find for the Gaussian part of the action
\begin{eqnarray}
	&&\frac{1}{2}\int_0^{\Lambda e^{-\ell}}\frac{\mathrm{d}^4 \mbf{q}^<}{(2\pi)^4}\left(r_0 + \left(\mbf{q}^{<}\right)^2 \right) |\mbf{M}^{<}(\mbf{q}^{<})|^2 \nonumber \\
	&& \qquad = \frac{1}{2}\int_0^{\Lambda} \frac{\mathrm{d}^4 \mbf{q}}{(2\pi)^4}e^{-6\ell}\zeta^2\left( e^{2\ell} r_0 + \mbf{q}^2 \right)|\mbf{M}(\mbf{q})|^2\,, \label{eq:rescaling}
\end{eqnarray}
and we choose $\zeta=e^{3\ell}$. The factor of $e^{2\ell}$ remaining in front of $r_0$ leads to the factor of $2$ appearing in the first term in Eqs.~(\ref{eq:quadratic_flow_1}) and (\ref{eq:quadratic_flow_2}). This number is referred to as the engineering dimension of $r_0$ and the fact that it is positive implies that $r_0$ is a relevant perturbation.

\subsection{RG Flow equations}

The term $\langle \mathcal{S}_{\rm int} \rangle_{>,0}$, with the appropriate momenta and field rescalings, yields the one-loop renormalization of the quadratic terms and the tree level renormalization of the quartic terms. Similarly, $\left\langle \mathcal{S}^{2}_{\rm int} \right\rangle_{>,0} - \left\langle \mathcal{S}_{\rm int} \right\rangle^2_{>,0}$ yields the one-loop renormalization of the quartic terms. A detailed presentation of the appropriate diagrams contributing to the flow equations is given in Appendix~\ref{app:diagrams}. Using the results presented there we find the flow equations
\begin{widetext}
\begin{eqnarray}
	\frac{\mathrm{d}\tilde{r}_{i,1}}{\mathrm{d}\ell} &=& 2\tilde{r}_{i,1} + 4\sum_{k=1}^{3}\left[ \frac{\lambda^{ik}_1}{1+ \tilde{r}_{k,1}} + \frac{\rho^{ik}}{1+\tilde{r}_{k,2}} \right] + 8\frac{\lambda_1^{ii}}{1+\tilde{r}_{i,1}} + 4 \frac{w^{ii}}{1+ \tilde{r}_{i,2}} \label{eq:quadratic_flow_1} \\
	\frac{\mathrm{d}\tilde{r}_{i,2}}{\mathrm{d}\ell} &=& 2\tilde{r}_{i,2} + 4\sum_{k=1}^{3}\left[ \frac{\lambda^{ik}_2}{1+ \tilde{r}_{k,2}} + \frac{\rho^{ik}}{1+\tilde{r}_{k,1}} \right] + 8\frac{\lambda_2^{ii}}{1+\tilde{r}_{i,2}} + 4 \frac{w^{ii}}{1+ \tilde{r}_{i,1}} \label{eq:quadratic_flow_2} \\
	\frac{\mathrm{d}\lambda^{ij}_1}{\mathrm{d}\ell} &=& -16 \frac{\lambda^{ij}_1 \lambda^{ji}_1}{(\tilde{r}_{i,1}+1)(\tilde{r}_{j,1} + 1)} - 8 \frac{\lambda^{ii}_1 \lambda^{ij}_1}{(\tilde{r}_{i,1}+1)^2} - 8 \frac{\lambda^{jj}_1 \lambda^{ji}_1}{(\tilde{r}_{j,1}+1)^2} - 4\sum_{k=1}^{3} \frac{\lambda^{ik}_1 \lambda^{kj}_1}{(\tilde{r}_{k,1}+1)^2} \nonumber \\
	&& -4\frac{\rho^{ji}w^{ii}}{(\tilde{r}_{i,2}+1)^2} -4\frac{\rho^{ij}w^{jj}}{(\tilde{r}_{j,2}+1)^2} - 4 \sum_{k=1}^{3} \frac{\rho^{ik}\rho^{jk}}{(\tilde{r}_{k,2}+1)^2} - 4 \frac{w^{ij}w^{ji}}{(\tilde{r}_{i,2}+1)(\tilde{r}_{j,2}+1)} \\
	\frac{\mathrm{d}\lambda^{ij}_2}{\mathrm{d}\ell} &=& -16 \frac{\lambda^{ij}_2 \lambda^{ji}_2}{(\tilde{r}_{i,2}+1)(\tilde{r}_{j,2} + 1)} - 8 \frac{\lambda^{ii}_2 \lambda^{ij}_2}{(\tilde{r}_{i,2}+1)^2} - 8 \frac{\lambda^{jj}_2 \lambda^{ji}_2}{(\tilde{r}_{j,2}+1)^2} - 4\sum_{k=1}^{3} \frac{\lambda^{ik}_2 \lambda^{kj}_2}{(\tilde{r}_{i,2}+1)^2} \nonumber \\
	&& -4\frac{\rho^{ji}w^{ii}}{(\tilde{r}_{i,1}+1)^2} -4\frac{\rho^{ij}w^{jj}}{(\tilde{r}_{j,1}+1)^2} - 4 \sum_{k=1}^{3} \frac{\rho^{ik}\rho^{jk}}{(\tilde{r}_{k,1}+1)^2} - 4 \frac{w^{ij}w^{ji}}{(\tilde{r}_{i,1}+1)(\tilde{r}_{j,1}+1)} \\
	\frac{\mathrm{d}\rho^{ij}}{\mathrm{d}\ell} &=& -8 \frac{\rho^{ij}\lambda^{ii}_1}{(\tilde{r}_{i,1}+1)^2} -8 \frac{\rho^{ij}\lambda^{jj}_2}{(\tilde{r}_{j,2}+1)^2} - 4\sum_{k=1}^{3}\left[ \frac{\lambda^{ik}_1\rho^{kj}}{(\tilde{r}_{k,1}+1)^2} + \frac{\rho^{ik}\lambda^{kj}_2}{(\tilde{r}_{k,2}+1)^2} \right] \nonumber \\ && -4 \frac{\lambda^{ij}_1 w^{jj}}{(\tilde{r}_{j,1}+1)^2} -4 \frac{\lambda^{ij}_2 w^{ii}}{(\tilde{r}_{i,2}+1)^2} - 16 \frac{\rho^{ij}\rho^{ij}}{(\tilde{r}_{i,1}+1)(\tilde{r}_{j,2}+1)} - 4 \frac{w^{ij}w^{ij}}{(\tilde{r}_{j,1}+1)(\tilde{r}_{i,2}+1)} \\
	\frac{\mathrm{d}w^{ij}}{\mathrm{d}\ell} &=& -8 \frac{w^{ij}\lambda_1^{ij}}{(\tilde{r}_{i,1}+1)(\tilde{r}_{j,1} +1)} -8 \frac{w^{ij}\lambda_2^{ij}}{(\tilde{r}_{i,2}+1)(\tilde{r}_{j,2} +1)} - 8 \frac{\rho^{ii}w^{ij}}{(\tilde{r}_{i,1} + 1)(\tilde{r}_{i,2} +1 )} - 8 \frac{\rho^{jj}w^{ij}}{(\tilde{r}_{j,1} + 1)(\tilde{r}_{j,2} +1 )} \nonumber \\
	&& -8\frac{\rho^{ij}w^{ij}}{(\tilde{r}_{i,1}+1)(\tilde{r}_{j,2}+1)} - 8\frac{\rho^{ij}w^{ij}}{(\tilde{r}_{j,1}+1)(\tilde{r}_{i,2}+1)} - 4\sum_{k=1}^{3} \frac{w^{ik}w^{kj}}{(\tilde{r}_{k,1}+1)(\tilde{r}_{k,2}+1)} \nonumber \\
	&& -4 \frac{w^{ji}w^{ii}}{(\tilde{r}_{i,1}+1)(\tilde{r}_{i,2}+1)} -4 \frac{w^{ij}w^{jj}}{(\tilde{r}_{j,1}+1)(\tilde{r}_{j,2}+1)}\,.\label{eq:quartic_flow_3}
\end{eqnarray}
\end{widetext}
Here we rescaled $\tilde{r}_{i,1}$ and $\tilde{r}_{i,2}$ by a factor of $\Lambda^2$ such that the ultraviolet cut-off is encountered when $\tilde{r}_{i,\mu}$ reaches unity. Note that $\tilde{r}_{x,1}=\tilde{r}_{y,2}$, $\tilde{r}_{x,2}=\tilde{r}_{y,1}$ and $\tilde{r}_{z,1}=\tilde{r}_{z,2}$ due to $C_4$ symmetry. In the following we will therefore only discuss $\tilde{r}_{i,1}$. The coupled non-linear differential equations~(\ref{eq:quadratic_flow_1})--(\ref{eq:quartic_flow_3}) can be solved numerically. However, prior to the study of the full solution it is helpful to first consider a number of limiting cases.

\subsection{Isotropic limit}

\begin{figure}
\centering
\includegraphics[width=0.9\columnwidth]{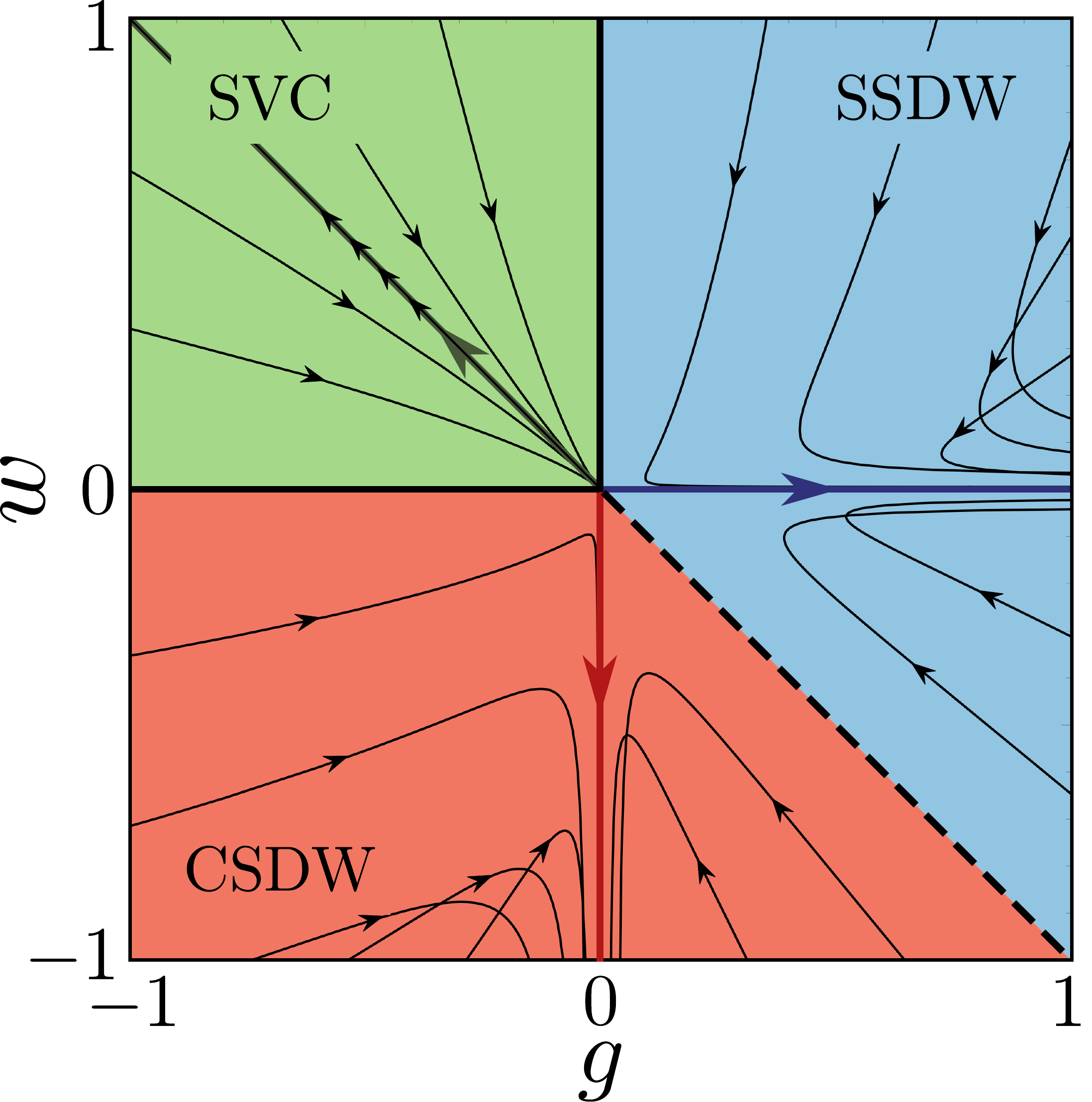}
\caption{\label{fig:isotropic_flow_diagram}(Color online) RG flow lines for the spin isotropic case (see Fig.~\ref{fig:mf_phase_diagram_w_points}) projected onto the $(g,w)$-plane. In all cases the coefficient $u$ flows towards negative infinity, as shown in Fig.~\ref{fig:specific_flows}. The behavior of the flow lines is understood in terms of the presence of three fixed trajectories. The stable parts of the fixed trajectories are here indicated by thick colored lines in respectively blue, red, and green. The crossing of flow lines is a consequence of the projection as the flows do not cross in $(u,g,w)$ space.}
\end{figure}

\begin{figure*}
\centering
\includegraphics[width=0.9\textwidth]{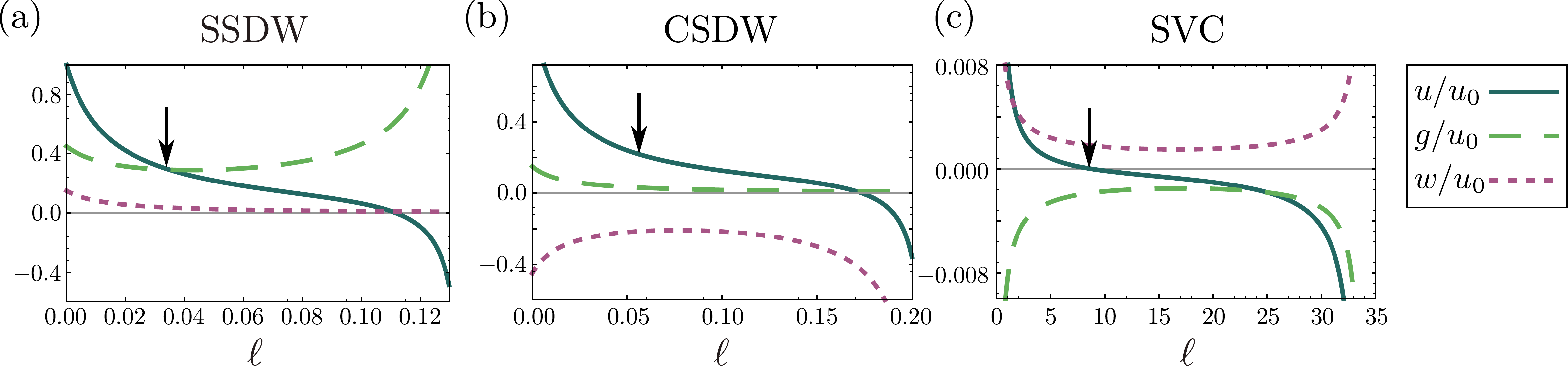}
\caption{\label{fig:specific_flows} (Color online) RG flows of the quartic coefficients for initial conditions corresponding to each of the three phases in the spin isotropic case. In (a) the initial conditions are $g_0/u_0=0.45$ and $w_0/u_0=0.15$ corresponding to the SSDW phase. In (b) we have $g_0/u_0=0.15$ and $w_0/u_0=-0.45$, within the CSDW phase. Finally, in (c) the initial conditions are $g_0/u_0=-0.45$ and $w_0/u_0=0.15$, in the SVC phase. In each case the flow of $u$ is towards negative infinity. This fact, combined with how the remaining two coefficients flow in each case implies that the transition is of first order. The points where the transitions become first-order are denoted by arrows and are (a) $u<g$, (b) $u<-w$, and (c) $u<0$.}
\end{figure*}

The isotropic limit of the above equations provides a good reference point for subsequent discussions. It can be achieved by having isotropic initial conditions for the quadratic coefficients: $\tilde{r}^{(0)}_{i,1}=\tilde{r}^{(0)}_{i,2}=r^{(0)}_0$. In this case there are only three coupled equations governing the flow of the quartic coefficients. In terms of the original coefficients we thus obtain
\begin{eqnarray}
	\dot{r}_0 &=& 2 r_0 + 16\frac{u}{r_0 + 1} - 4\frac{g}{r_0 +1} + 4 \frac{w}{r_0 + 1} \label{eq:r_isotropic} \\
	\dot{u} &=& -28\frac{u^2}{(r_0 + 1)^2} - 8 \frac{g^2}{(r_0+1)^2} + 8\frac{ug}{(r_0+1)^2} \nonumber \\ && \qquad - 8\frac{uw}{(r_0+1)^2} -8 \frac{w^2}{(r_0+1)^2} \\
	\dot{g} &=& 20\frac{g^2}{(r_0 + 1)^2} - 24 \frac{ug}{(r_0+1)^2} + 8\frac{gw}{(r_0+1)^2} \\
	\dot{w} &=& -20\frac{w^2}{(r_0+1)^2} - 24 \frac{uw}{(r_0+1)^2} -8\frac{gw}{(r_0+1)^2}\,, \label{eq:w_isotropic}
\end{eqnarray}
where the dot denotes differentiation with respect to $\ell$. Various aspects of these equations have been studied previously, see Refs.~\onlinecite{qi09,millis10,batista11,fernandes12}. Here, we will focus on the fixed trajectories. Numerical solutions of Eqs.~(\ref{eq:r_isotropic})--(\ref{eq:w_isotropic}) are presented in Figs.~\ref{fig:isotropic_flow_diagram} and \ref{fig:specific_flows}, and below we discuss the properties of these equations.
We note in passing the well-known phenomena that fluctuations serve to suppress the value of $r_0$ for which a transition occurs. In the absence of fluctuations, i.e. with no quartic terms in Eq.~(\ref{eq:r_isotropic}), the transition occurs at $r_0 = 0$, signalled by the fact that for the initial condition $r^{(0)}_0 = 0$ the flow equations yield $\dot{r}_0=0$. However, when quartic (or higher) terms are present, the value of $r_0^{(0)}$ for which $\dot{r}_0=0$ is shifted downwards. In the discussion below we will assume that we are at the magnetic transition, $r_0^{(0)}=r^c$, such that $\dot{r}_0=0$. Then we can ignore the flow of $r_0$, and absorb a factor of $(r^c+1)^{-2}$ in the flow parameter $\ell$. We do not seek to determine the value of $r^{c}$ here.

\subsubsection{Fixed trajectories}

In this case the Gaussian fixed point is unstable except for flows with fine-tuned initial conditions which we discuss below. The RG flows are instead governed by a number of fixed trajectories. These are trajectories for which the ratio of two coefficients tends to a constant although the coefficients themselves might diverge at a finite value of $\ell=\ell_c$. These fixed trajectories are
\begin{eqnarray}
	\text{SSDW:}& \qquad \left(\frac{w}{g},\frac{u}{g}\right)^{\ast} &= (0,-1)\,,\\
	\text{CSDW:}& \qquad \left(\frac{g}{w},\frac{u}{w}\right)^{\ast} &= (0,1)\,,\\
	\text{SVC:}& \qquad \left(\frac{w}{g},\frac{g}{u}\right)^{\ast} &= (-1,0)\,.
\end{eqnarray}
For each fixed trajectory there are associated basins of attraction and basins of repulsion. In these regions the flows are either attracted or repelled by the fixed trajectories. Focusing first on the $(g,w)$--plane we consider each of the fixed trajectories in turn.

A straightforward stability analysis reveals that the fixed trajectory identified with the SSDW phase is indeed attractive for $g>0$ and repulsive for $g<0$. The $g>0$ branch of the fixed trajectory thus acts to ensure that the flows within the SSDW region are attracted to the $(w/g)^{\ast}=0$ line. This is the reason for associating the fixed trajectory with the SSDW phase, the stable part is depicted by a dark blue line in Fig.~\ref{fig:isotropic_flow_diagram}. In contrast, the $g<0$ branch acts as a separatrix between the SVC and CSDW phase, and coincides with the mean-field phase boundary between these two phases. This implies that any accidental degeneracies between the two are avoided. By this, we mean that even if the bare interaction parameters $g$ and $w$, as derived e.g. from a microscopic band structure calculation, are such that the system is located close to a phase boundary, fluctuations will inevitably renormalize these parameters. As a result, the RG flow will bring the system away from the near degeneracy, deep into the magnetic phase it started in. Similarly, the fixed trajectory $(g/w)^{\ast}=0$ is attractive for $w<0$ and repulsive for $w>0$. The repulsive branch coincides with the mean-field phase boundary between the SSDW and SVC phases, and forms a separatrix between the two. As previously, this prevents any accidental near degeneracies from occuring. On the other hand, the attractive branch, $w<0$, lies deep within the CSDW phase, and is indicated by a dark red line in Fig.~\ref{fig:isotropic_flow_diagram}. Finally, the fixed trajectory associated with the SVC phase, $(w/g)^{\ast}=-1$, is attractive for $w>0$ and $g<0$, and is shown by a dark green line in Fig.~\ref{fig:isotropic_flow_diagram}. Hence, flow lines originally within the SVC phase also remain within the phase. The trajectory is repulsive for $g>0$ and $w<0$, and in this case it acts as a separatrix between the CSDW and SSDW phases, coinciding with the mean-field phase boundary between the two. Thus, attractive and repulsive branches of the fixed trajectories act in tandem to prevent the occurence of any accidental near degeneracies in the system. This implies that the mean-field phase diagram is stable against fluctuations. The behavior described can be seen in Fig.~\ref{fig:isotropic_flow_diagram}.

The fixed trajectories involving $u$ serve a different purpose. A stability analysis similar to the one above shows that for the SSDW phase the fixed trajectory $(u/g)^{\ast}=-1$ is stable for $g>0$ and $u<0$. Combining this with the fact that $u(\ell \rightarrow \ell_c) \rightarrow -\infty$, as shown in Fig.~\ref{fig:specific_flows}, implies that the magnetic transition is driven first order by the fluctuations. This is seen from the fact that the action becomes unbounded under the RG flow. Formally this would require the introduction of higher-order terms in the action to ensure that it remains bounded. Here we will follow standard procedure and assume that higher-order terms exist such that the action is bounded and interpret the negative quartic term as a signal of a first-order transition. Note that the transition turns first order when $u<g$ which happens for $\ell < \ell_c$. Similar arguments hold for $(u/w)^{\ast}=1$ in the CSDW phase. This trajectory is stable for $u<0$ and $w<0$. In this case the transition becomes first order when $u<-w$. Again, this occurs for $\ell < \ell_c$. In the SVC case, the fixed trajectory involving $u$ is $(g/u)^{\ast}=0$. This is stable within the SVC phase, $g<0$ and $w>0$, for $u<0$. As above this implies a first-order transition, which occurs when $u<0$ and $\ell < \ell_c$. The values of $\ell$ for which the transitions turn first order are denoted by arrows in Fig.~\ref{fig:specific_flows}.

\subsubsection{Gaussian fixed point}

Finally, we briefly comment on the fate of the Gaussian fixed point in the isotropic case. As explained above, the RG flows are governed by the fixed trajectories and will approach the stable branches asymptotically as $\ell \rightarrow \ell_c$. However, a different behavior emerges if the initial conditions, $g_{(0)}$ and $w_{(0)}$, lie on one of the trajectories $(w/g)^{\ast}=0$, $(g/w)^{\ast}=0$ or $(w/g)^{\ast}=-1$. If the initial conditions are on one of the unstable branches of the fixed trajectories, the flow is towards the Gaussian fixed point, i.e. $u=g=w=0$. This occurs regardless of the initial condition for $u$, as long as the free energy is initially bounded. Note that for initial conditions on the unstable branches, two of the magnetic phases are accidentally degenerate. It is unlikely that such a scenario would occur in realistic systems however. Even if the initial conditions could be fine-tuned, any infinitesimal perturbation would displace the flow from the unstable branch of the fixed trajectory. For initial conditions on the stable branches of the fixed trajectories, there is no accidental degeneracy. In these cases the flows are directed along the fixed trajectories towards a first-order transition, and away from the Gaussian fixed point.

\subsection{Strongly anisotropic limits}

The strongly anisotropic cases can be studied in a similar manner. Here we review our results presented in Ref.~\onlinecite{christensen17b} Let us commence with the case $\tilde{r}_{x,1} \ll \tilde{r}_{y,1},\tilde{r}_{z,1}$, corresponding to $\alpha_1 \ll \alpha_2,\alpha_3$. The bare free energy in this case is given in Eq.~(\ref{eq:free_energy_alpha1}). Importantly, we find that no additional terms arise as a result of the RG flow, as the $M_{x,1}$ and $M_{y,2}$ modes remain decoupled from the rest. In terms of the coefficients of Eq.~(\ref{eq:free_energy_alpha1}) the relevant flow equations for $\alpha_1 \ll \alpha_{2}, \alpha_3$ are
\begin{eqnarray}
	\dot{\tilde{r}}_{x,1} &=& 2\tilde{r}_{x,1} + \frac{8u_{\alpha_1}}{\tilde{r}_{x,1}+1} -\frac{4g_{\alpha_1}}{\tilde{r}_{x,1}+1} \label{eq:r11_alpha1} \\
	\dot{u}_{\alpha_1} &=& - \frac{20u_{\alpha_1}^2}{(\tilde{r}_{x,1}+1)^2} - \frac{8g_{\alpha_1}^2}{(\tilde{r}_{x,1}+1)^2} +  \frac{8u_{\alpha_1}g_{\alpha_1}}{(\tilde{r}_{x,1}+1)^2} \label{eq:u_alpha1} \\
	\dot{g}_{\alpha_1} &=& - \frac{24u_{\alpha_1}g_{\alpha_1}}{(\tilde{r}_{x,1}+1)^2} +  \frac{12g_{\alpha_1}^2}{(\tilde{r}_{x,1}+1)^2}\,, \label{eq:g_alpha1}
\end{eqnarray}
where $u_{\alpha_1}=\rho^{xy} + \lambda_1^{xx}$ and $g_{\alpha_1}=\rho^{xy}-\lambda^{xx}_1$. Note that this forms a closed set of equations signalling a decoupling of the order parameters. The flow equations governing the remaining spin components are zero due to the initial conditions imposed by the bare free energy Eq.~(\ref{eq:free_energy_alpha1}). Hence, no additional terms are generated under the RG flow. Crucially, for Eqs.~(\ref{eq:r11_alpha1})--(\ref{eq:g_alpha1}) the Gaussian fixed point, $u_{\alpha_1}^{\ast}=g_{\alpha_1}^{\ast}=0$, is stable for a range of initial conditions. This implies an enhanced degeneracy between the magnetic states, because $C_2$ and $C_4$ symmetric states cannot be distinguished in the absence of quartic terms in the free energy. To split them, higher order terms are required.

To see the origin of the stable Gaussian fixed point, let us assume that we are right at the magnetic transition, $\tilde{r}_{x,1}^{(0)} = \tilde{r}^{c}_{x,1}$ such that $\dot{\tilde{r}}_{x,1}=0$, and consider the fixed trajectories of Eqs.~(\ref{eq:u_alpha1}) and (\ref{eq:g_alpha1}). These are
\begin{eqnarray}
	\left(\frac{u_{\alpha_1}}{g_{\alpha_1}}\right)^{\ast} = 2\,, \qquad \left(\frac{u_{\alpha_1}}{g_{\alpha_1}}\right)^{\ast}=-1\,.
\end{eqnarray}
A stability analysis shows that $(u_{\alpha_1}/g_{\alpha_1})^{\ast}=2$ is repulsive for $g_{\alpha_1} > 0$ while $(u_{\alpha_1}/g_{\alpha_1})^{\ast}=-1$ is repulsive for $g_{\alpha_1}<0$. Hence, for $u^{(0)}_{\alpha_1}>0$, flows within the fan formed by $(u_{\alpha_1}/g_{\alpha_1})^{\ast}=2$ and $(u_{\alpha_1}/g_{\alpha_1})^{\ast}=-1$ can only flow to the Gaussian fixed point. This is similar to the $N=1$ case studied in Ref.~\onlinecite{fernandes12}.

The case with $\tilde{r}_{z,1}\ll\tilde{r}_{x,1},\tilde{r}_{y,1}$, i.e. $\alpha_3 \ll \alpha_1,\alpha_2$, behaves in a similar manner. Here, the bare action is given in Eq.~(\ref{eq:free_energy_alpha3}) and, as previously, no additional terms are generated under the one-loop RG flow i.e. the equations governing $M_{z,1}$ and $M_{z,2}$ decouple from the rest. Taking the appropriate limit of Eqs.~(\ref{eq:quadratic_flow_1})--(\ref{eq:quartic_flow_3}) we find, for $\alpha_3 \ll \alpha_1, \alpha_2$,
\begin{eqnarray}
	\dot{\tilde{r}}_{z,1} &=& 2\tilde{r}_{z,1} + 8\frac{u_{\alpha_3} + w_{\alpha_3}}{\tilde{r}_{z,1}+1} -4\frac{g_{\alpha_3}+w_{\alpha_3}}{\tilde{r}_{z,1}+1} \label{eq:r11_alpha3} \\
	\dot{u}_{\alpha_3} + \dot{w}_{\alpha_3} &=& -20 \frac{(u_{\alpha_3} + w_{\alpha_3})^2}{(\tilde{r}_{z,1}+1)^2} - 8 \frac{(g_{\alpha_3} + w_{\alpha_3})^2}{(\tilde{r}_{z,1}+1)^2} \nonumber \\ && \qquad  + 8 \frac{(u_{\alpha_3}+w_{\alpha_3})(g_{\alpha_3}+w_{\alpha_3})}{(\tilde{r}_{z,1}+1)^2} \label{eq:u_alpha3} \\
	\dot{g}_{\alpha_3} + \dot{w}_{\alpha_3} &=& -24 \frac{(u_{\alpha_3} + w_{\alpha_3})(g_{\alpha_3}+w_{\alpha_3})}{(\tilde{r}_{z,1}+1)^2} \nonumber \\ && \qquad + 12 \frac{(g_{\alpha_3}+w_{\alpha_3})^2}{(\tilde{r}_{z,1}+1)^2}\,. \label{eq:g_alpha3}
\end{eqnarray}
The specific forms of the equations have been chosen to highlight the fact that there are only two independent quartic coefficients in this case, just as for the $\alpha_1\ll\alpha_2,\alpha_3$ case above. Here, however, $u_{\alpha_3}=\rho^{zz}+\lambda_1^{zz}$, $g_{\alpha_3}=\rho^{zz}-\lambda_1^{zz}$ and $w_{\alpha_3}=w^{zz}$. Note that Eqs.~(\ref{eq:r11_alpha3})--(\ref{eq:g_alpha3}) have identical structure to Eqs.~(\ref{eq:r11_alpha1})--(\ref{eq:g_alpha1}), yet with different coupling constants. Hence, the results of the previous analysis carry over and the Gaussian fixed point, $(u_{\alpha_3}+w_{\alpha_3})^{\ast}=(g_{\alpha_3}+w_{\alpha_3})^{\ast}=0$, is stable for a range of initial conditions. The fixed trajectories (assuming $\dot{\tilde{r}}_{z,1}=0$) are
\begin{eqnarray}
	\left(\frac{u_{\alpha_3}+w_{\alpha_3}}{g_{\alpha_3}+w_{\alpha_3}}\right)^{\ast}=2\,, \qquad \left(\frac{u_{\alpha_3}+w_{\alpha_3}}{g_{\alpha_3}+w_{\alpha_3}}\right)^{\ast}=-1\,.
\end{eqnarray}
As previously, the fixed trajectories form a fan of repulsive separatrices for $u^{(0)}_{\alpha_3}+w^{(0)}_{\alpha_3}>0$, implying that flows within this fan terminate at the Gaussian fixed point.

\begin{figure}
\centering
\includegraphics[width=0.9\columnwidth]{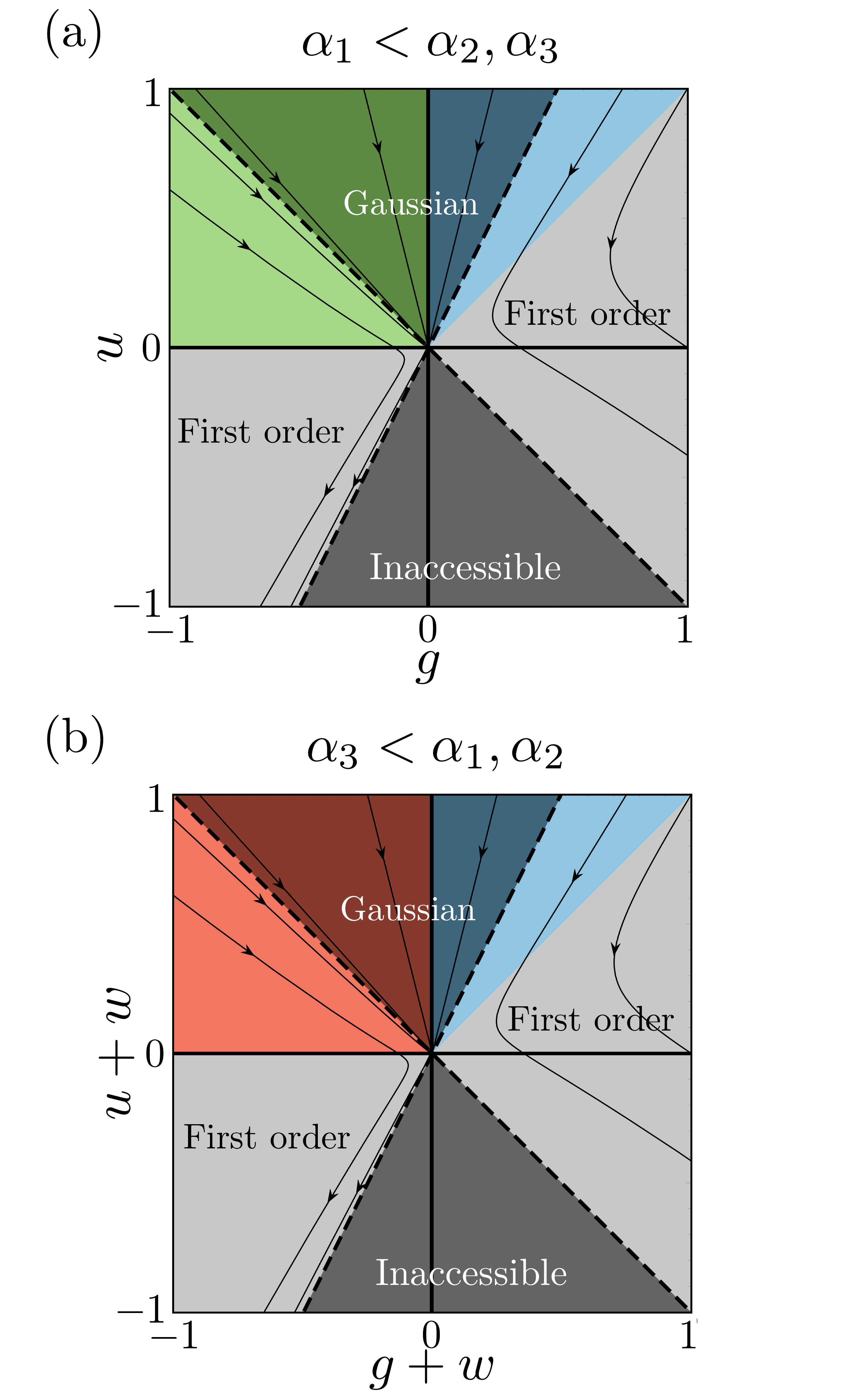}
\caption{\label{fig:large_anis_flow}(Color online) Flow diagrams for the cases (a) $\alpha_1 < \alpha_2,\alpha_3$ and (b) $\alpha_3 < \alpha_1,\alpha_2$. Dark areas denote the Gaussian basin of attraction. Lighter areas denote regions where the free energy is bounded, but the flow leads to a first order transition. The free energy is unbounded within the gray areas. The dark gray area is entirely inaccessible to flows originating outside this region, owing to the separatrices given by the fixed trajectories.}
\end{figure}

We summarize the above findings in Fig.~\ref{fig:large_anis_flow}. The basin of attraction associated with the Gaussian fixed point (dark shaded region) is seen to occupy a large region of the flow diagrams. This demonstrates that this fixed point will dominate the flows for a significant range of initial conditions. Note that the case $\alpha_2\ll\alpha_1,\alpha_3$ is similar to $\alpha_1\ll\alpha_2,\alpha_3$.

\subsection{General anisotropic case}\label{sec:flows_details}

\begin{figure*}
\centering
\includegraphics[width=0.9\textwidth]{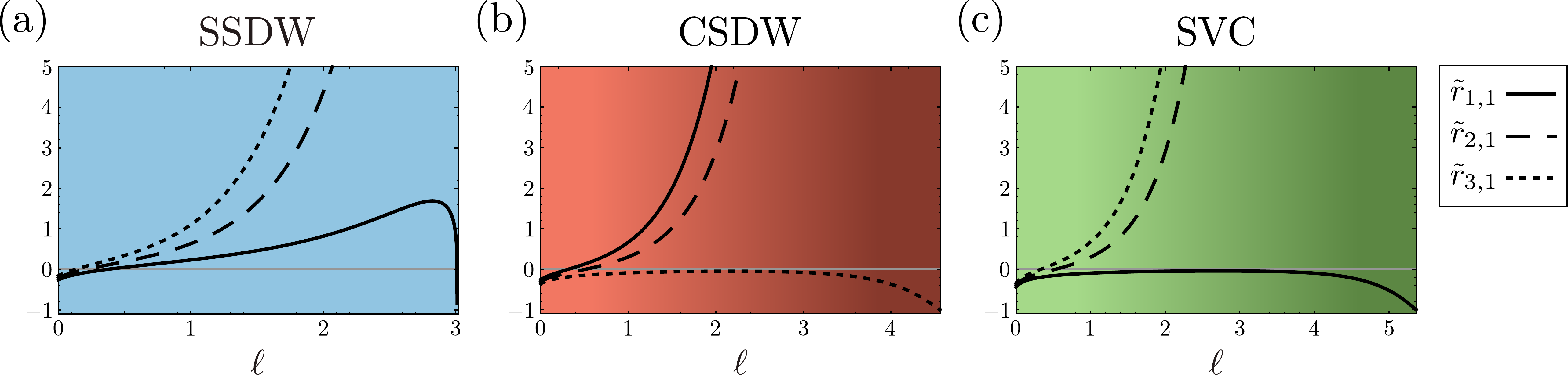}
\caption{\label{fig:anisotropic_quadratic_flows} (Color online) RG flows of the quadratic coefficients $\tilde{r}_{i,1}$ (the flow of $\tilde{r}_{i,2}$ follows from $\tilde{r}_{\bar{i},2} = \tilde{r}_{i,1}$), in the presence of SOC for initial quartic coefficients corresponding to either of the three phases, (a) SSDW, (b) CSDW, and (c) SVC. A decoupling is seen to occur in all cases, with the smallest of the $\alpha$s determining the components that condense. Similar decouplings are observed regardless of which $\alpha$ is chosen to be the smallest. The darker shading in (b) and (c) denotes the approach of the Gaussian fixed point. In (a) we used $\tilde{r}^{(0)}_{x,1}=-0.26137$, $(\tilde{r}^{(0)}_{y,1}-\tilde{r}^{(0)}_{x,1})/u_{(0)}=0.05$, and $(\tilde{r}^{(0)}_{z,1}-\tilde{r}^{(0)}_{x,1})/u_{(0)}=0.1$ with $g_{(0)}/u_{(0)}=0.6$ and $w_{(0)}/u_{(0)}=0.8$. In (b) $\tilde{r}^{(0)}_{z,1}=-0.35639$, $(\tilde{r}^{(0)}_{y,1}-\tilde{r}^{(0)}_{z,1})/u_{(0)}=0.05$, and $(\tilde{r}^{(0)}_{x,1}-\tilde{r}^{(0)}_{z,1})/u_{(0)}=0.1$ with $g_{(0)}/u_{(0)}=0.3$ and $w_{(0)}/u_{(0)}=-0.4$. In (c) $\tilde{r}^{(0)}_{x,1}=-0.44859$, $(\tilde{r}^{(0)}_{y,1}-\tilde{r}^{(0)}_{x,1})/u_{(0)}=0.05$ and $(\tilde{r}^{(0)}_{z,1}-\tilde{r}^{(0)}_{x,1})/u_{(0)}=0.1$ with $g_{(0)}/u_{(0)}=-0.4$ and $w_{(0)}/u_{(0)}=0.6$.}
\end{figure*}

The case where only one $\alpha$-coefficient is dominant seems contrived given that SOC is small and of the order $\sim 10$~meV~\cite{borisenko16} in the iron pnictides. Nevertheless, as we show now, the universal behavior of the system at weaker anisotropies is the same as the case of stronger anisotropy. This happens due to the fact that anisotropy-inducing terms constitute RG relevant perturbations.

Here we consider the full numerical solution of the general RG equations~(\ref{eq:quadratic_flow_1})--(\ref{eq:quartic_flow_3}), including anisotropies. We start by considering the flow of the quadratic coefficients and note that an initial small splitting between the $\tilde{r}_i$ rapidly grows under the RG flow, signalling a decoupling of the spin components. To be explicit, we consider a variety of initial conditions for the quartic coefficients, starting in each of the three phases, SSDW, CSDW, and SVC. The flows of the mass terms are depicted in Fig.~\ref{fig:anisotropic_quadratic_flows} and a decoupling is seen to occur regardless of initial conditions.
Despite a minute difference in the bare values of the $\tilde{r}_i$s, the splitting grows to become substantial under the RG flow. As previously, the bare value of the smallest $\tilde{r}^{(0)}_i$ can in principle be chosen such that $\dot{\tilde{r}}_i=0$. However, finding the precise numerical value of $\tilde{r}^c$ can be challenging since the flows of the quadratic coefficients are coupled. Hence, the flow of the quadratic coefficients will be as depicted in Fig.~\ref{fig:anisotropic_quadratic_flows}: two coefficients exhibit runaway flows, $\tilde{r}_{j\neq i}(\ell \rightarrow \ell_c) \rightarrow \infty$. The spin components associated with these coefficients acquire asymptotically infinite masses under the RG flow and cannot condense. On the other hand, the smallest coefficient $\tilde{r}_{i}(\ell \rightarrow \ell_c) \rightarrow -1$ signaling a condensation of the associated degrees of freedom. We note that due to the rescaling of $\tilde{r}_i$ by $\Lambda^2$ the value $-1$ denotes the cutoff scale [c.f. the discussion following Eq.~(\ref{eq:second_class_integral})]. The actual value of $\tilde{r}^{c}$ is located between $\tilde{r}_i^{(0)}$ and the $\tilde{r}_{j \neq i}^{(0)}$. In the anisotropic cases considered here we typically determined the value of $\tilde{r}^c$ to six significant digits, allowing us to follow the flow to rather large values of $\ell$ before $\ell_c$ is encountered, which in turn makes the decoupling more apparent.

\begin{figure}
\centering
\includegraphics[width=\columnwidth]{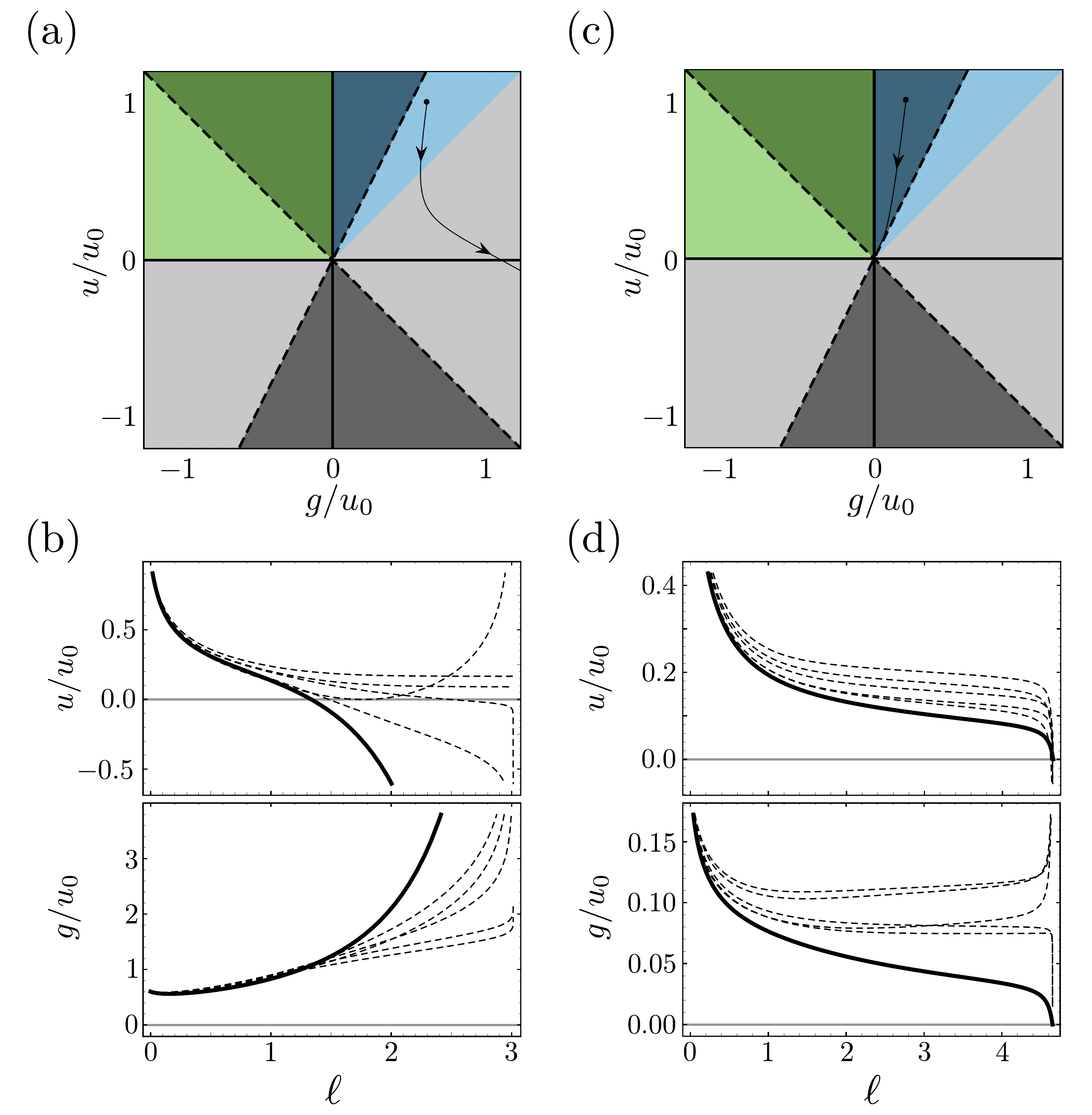}
\caption{\label{fig:finite_alpha_flows_ssdw}(Color online) RG flows of the quartic coefficients for $\alpha_1 < \alpha_2,\alpha_3$. In (a) and (b) initial conditions are $(\rho^{i\bar{j}}+\lambda^{ij}_1)_{(0)}=u_{(0)}=0.1$ and $(\rho^{i\bar{j}}+\lambda^{ij}_1)_{(0)}=g_{(0)}=0.06$, outside the Gaussian region. The magnitude of the spin anisotropy is chosen as in Fig.~\ref{fig:anisotropic_quadratic_flows}. In (a) the flow of the quartic coefficients $\rho^{xy}+\lambda_1^{xx}=u_{\alpha_1}$ and $\rho^{xy}-\lambda_1^{xx}=g_{\alpha_1}$ is depicted. In (b) the flows of all the quartic coefficients, $u^{ij}$ and $g^{ij}$, are shown. The flows of the quartic coefficients governing the spin components with diverging masses are denoted by dashed lines, while the flow of $u_{\alpha_1}$ and $g_{\alpha_1}$ are highlighted in black. For the case in (c) and (d) the initial conditions were chosen to be $u_{(0)}=0.1$ and $g_{(0)}=0.02$, inside the Gaussian region. The flows of $u_{\alpha_1}$ and $g_{\alpha_1}$ clearly terminate at the Gaussian fixed point. The magnitude of the spin anisotropy is the same as in (a) and (b).}
\end{figure}
\begin{figure}
\centering
\includegraphics[width=\columnwidth]{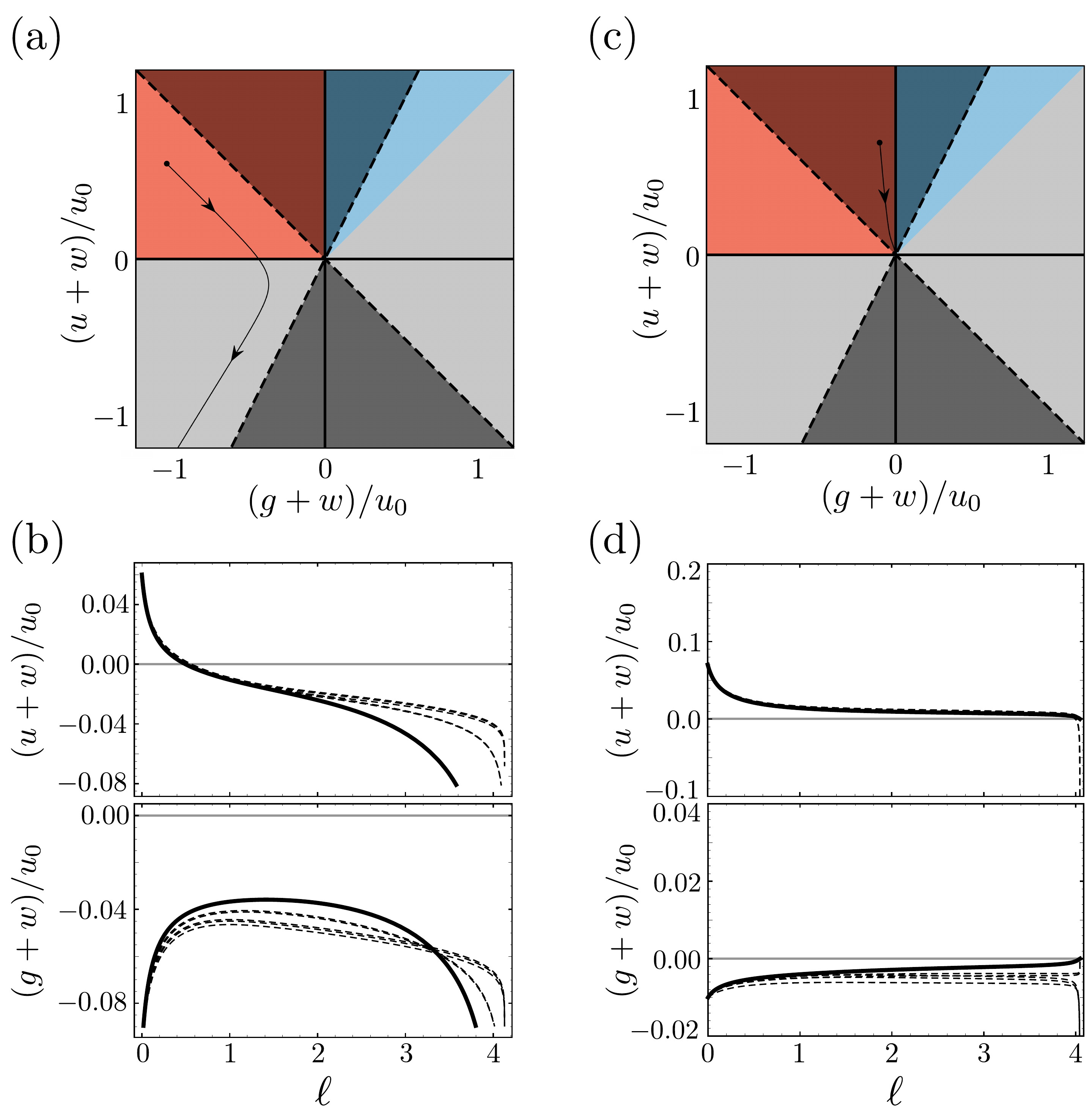}
\caption{\label{fig:finite_alpha_flows_csdw}(Color online) RG flows for the quartic coefficients in the case where $\alpha_3 < \alpha_1,\alpha_2$. The magnitude of the spin anisotropy is as in Figs.~\ref{fig:anisotropic_quadratic_flows} and \ref{fig:finite_alpha_flows_ssdw}. As in Fig.~\ref{fig:finite_alpha_flows_ssdw} the initial conditions in (a) and (b) were chosen outside the Gaussian region, $(\rho^{i\bar{j}}+\lambda^{ij}_1+w^{ij})_{(0)}=(u+w)_{(0)}=0.06$ and $(\rho^{i\bar{j}}-\lambda^{ij}_1+w^{ij})_{(0)}=(g+w)_{(0)}=-0.1$ (we take $u_0=0.1$). In (a) the flow of the quartic coefficients $(u+w)^{zz}=u_{\alpha_3}+w_{\alpha_3}$ and $(g+w)^{zz}=g_{\alpha_3}+w_{\alpha_3}$ is depicted. In (b) the flows of the coefficients $(u+w)^{ij}$ and $(g+w)^{ij}$ are shown, with $u_{\alpha_3}+w_{\alpha_3}$ and $g_{\alpha_3}+w_{\alpha_3}$ highlighted in black. The dashed lines denote the flows of the quartic coefficients associated with spin components with diverging quadratic terms. In (c) and (d) the initial conditions are $(u+w)_{(0)}=0.07$ and $(g+w)_{(0)}=-0.01$, which lie inside the Gaussian basin of attraction. In this case the flows of $u_{\alpha_3}+w_{\alpha_3}$ and $g_{\alpha_3}+w_{\alpha_3}$ terminate at the Gaussian fixed point. The spin anisotropy is the same as in (a) and (b).}
\end{figure}

The above decoupling is a consequence of the RG relevance of the quadratic terms, as the presence of spin anisotropy reduces the symmetry of the model from $O(3)\times O(3)$ in the isotropic case, to $\mathbb{Z}_2 \times \mathbb{Z}_2$ in the anisotropic case. This implies that the flow diagrams of Fig.~\ref{fig:large_anis_flow} are expected to capture the salient features of the system. This is confirmed by considering the flows of the quartic coefficients.

In Fig.~\ref{fig:finite_alpha_flows_ssdw} we depict the flows of all the quartic coefficients in the case where $\alpha_1 < \alpha_2,\alpha_3$.
Guided by the strongly anisotropic case of Fig.~\ref{fig:large_anis_flow}(a) we consider bare values both within and outside the Gaussian basin of attraction and confirm that the quartic coefficients indeed exhibit two distinct flows. To facilitate comparison with the strongly anisotropic case we consider the combinations $u^{ij}=\rho^{i\bar{j}}+\lambda^{ij}_1$ and $g^{ij}=\rho^{i\bar{j}}-\lambda_1^{ij}$. Here ${\bar{i}}=y,x,z$, as previously. In Figs.~\ref{fig:finite_alpha_flows_ssdw}(a) and (c) we show the flow of $u^{xx} \equiv u_{\alpha_1}$ and $g^{xx} \equiv g_{\alpha_1}$ in the phase diagram of the strongly anisotropic case to further illustrate how the behavior depends on the choice of bare values. For completeness we include the flows of $u^{ij}$ and $g^{ij}$  in Figs.~\ref{fig:finite_alpha_flows_ssdw}(b) and (d), with $u_{\alpha_1}$ and $g_{\alpha_1}$ highlighted in black. Note that the flow of the remaining components (dashed lines) do not affect the magnetic order, as these govern spin components with asymptotically infinite masses.

Similarly, in Fig.~\ref{fig:finite_alpha_flows_csdw} we consider the case with $\alpha_3<\alpha_1,\alpha_2$. 
In this case we choose bare quartic values guided by Fig.~\ref{fig:large_anis_flow}(b). Hence we consider $(u+w)^{ij}=\rho^{i\bar{j}}+\lambda_1^{ij}+w^{ij}$ and $(g+w)^{ij}=\rho^{i\bar{j}}-\lambda_1^{ij}+w^{ij}$. In Figs.~\ref{fig:finite_alpha_flows_csdw}(a) and (c) we show the flows of $(u+w)^{zz} \equiv u_{\alpha_3}+w_{\alpha_3}$ and $(g+w)^{zz} \equiv g_{\alpha_3} + w_{\alpha_3}$. As expected, the Gaussian fixed point is attractive for a range of bare values, as seen in Fig.~\ref{fig:finite_alpha_flows_csdw}(c). In Figs.~\ref{fig:finite_alpha_flows_csdw}(b) and (d) we show the flows of all $(u+w)^{ij}$ and $(g+w)^{ij}$. As before the flow of the relevant quartic coefficiens, $u_{\alpha_3}+w_{\alpha_3}$ and $g_{\alpha_3}+w_{\alpha_3}$, that describe the interactions among the components that condense,  is highlighted in black. The remaining components, governing the infinitely massive spin components, are depicted by dashed lines.

In general we find that the strongly anisotropic flow diagrams of Fig.~\ref{fig:large_anis_flow} provide a good description of the system. However, to ensure that the decoupling between the two sectors occurs properly, the value of the smallest $\tilde{r}_{i}$ must be close to $\tilde{r}^c$. Otherwise, the flows will terminate before the masses of the remaining spin components can approach infinity and a non-zero coupling will still exist between the two sectors. In such cases the flow can leave the Gaussian basin of attraction due to the influence of the more massive spin components. A similar phenomenon can occur if the bare values are very close to the boundary of the Gaussian basin of attraction and the differences between the $\tilde{r}_i$ are chosen to be very small compared to the distance to this boundary. In this case, the flow can leave the Gaussian region before a decoupling of the order parameters occurs, and the quartic coefficients will instead flow towards one of the fixed trajectories. Hence, the boundaries of the Gaussian basins of attraction in Figs.~\ref{fig:finite_alpha_flows_ssdw} and \ref{fig:finite_alpha_flows_csdw} are not exact but depend on the relative values of $\alpha_1$, $\alpha_2$ and $\alpha_3$, along with the bare values of the quartic coefficients.

\section{Conclusions and Outlook}\label{sec:conclusions}

We considered the impact of SOC on the magnetic phase diagrams of the iron pnictides. The presence of a sizeable SOC is attested both by direct measurements~\cite{borisenko16} and by the observation of a substantial spin anisotropy in a variety of experimental probes~\cite{lipscombe10,qureshi12,wang13,song13,steffens13,li11,hirano12,curro17,
wasser15,allred16a}. SOC breaks the spin rotational invariance which manifests itself by the appearance of spin anisotropic quadratic terms in the action. At the mean-field level this can lead to frustration between the quadratic and quartic coefficients, when the latter selects $C_4$ phases. Near $T_{\rm mag}$ the quadratic coefficients lift the frustration, while at lower temperatures new phases appear in order to balance the effects of the quadratic and quartic coefficients. These phases are superpositions of the well-known SSDW, CSDW and SVC phases. 

Going beyond mean-field theoy, we took into account the effect of magnetic fluctuations by adopting an RG approach. To ensure that the RG procedure is controlled we considered only $T=0$. In the absence of SOC, magnetic fluctuations leave the mean-field phase boundaries intact. In fact, we find that the mean-field phases become more robust; the stable parts of the fixed trajectories are located deep within each phase. Additionally, the magnetic quantum phase transitions become first order under the RG flow. This is in stark contrast to the situation when SOC is included and the quadratic terms become anisotropic. Since the anisotropies are relevant under the RG flow we find a rapid decoupling of the spin components. Thus, only a subset of these condense at the magnetic transition. This has several important consequences. It leads to a basin of attraction for the stable Gaussian fixed point, which covers a large range of initial conditions. This implies that the Gaussian fixed point can play a role for a wide range of initial conditions and is thus less sensitive to details of the system. The Gaussian fixed point results in an enhanced degeneracy between the $C_2$ and $C_4$ magnetic states and higher-order coefficients are required to break this degeneracy. Furthermore, depending on which SOC-induced coefficient $\alpha_i$ is the smallest, one of the $C_4$ phases is ruled out, implying a connection between the direction of the spin anisotropy (in-plane or out-of-plane) and the type of $C_4$ order possible.

Most importantly, these results provide compelling evidence that the proliferation of magnetic phases in the vicinity of the putative magnetic QCP is due to the interplay between SOC and magnetic fluctuations. In Ba$_{1-x}$Na$_x$Fe$_2$As$_2$ such a proliferation of magnetic phases was recently observed~\cite{wang16}. Here a host of additional magnetic phases appear as the putative QCP is approached. Within our model, it is natural to expect additional phases to appear close to the putative QCP due to the enhanced magnetic degeneracy arising as a consequence of the spin anisotropy. We note that the low-temperature mean-field results of e.g. Refs.~\onlinecite{gastiasoro15,christensen17} both predict values of $g$ and $w$ within the Gaussian basin of attraction, and thus our results are relevant for these models.

Additionally, the frustration provides an explanation of why, in most iron-based superconductors, the magnetic $C_4$ phase only appears inside the magnetic $C_2$ phase~\cite{avci14a,wang16,hassinger,bohmer15a,
khalyavin14,allred15a,taddei16a,taddei17a}. This is most clear in the results presented in Fig.~\ref{fig:case_a}. With the change of an external parameter, such as doping or pressure, the quartic coefficients change from favoring a $C_2$ phase to favoring a $C_4$ phase. However, as evidenced from the fact that the magnetic moments are in-plane and parallel to the ordering vector, $\alpha_1 < \alpha_2,\alpha_3$. Thus, the leading instability will be a $C_2$ SSDW phase, and only at lower temperatures will the $C_4$ (CSDW) phase set in. Depending on the ratio between the spin anisotropic coefficients, our results show that an additional phase mixing the $C_2$ and $C_4$ phases may be realized as well. Recent experiments in Ba$_{1-x}$Na$_x$Fe$_2$As$_2$ indicated that other phases may emerge near the transition from $C_2$ to $C_4$ magnetism. It would be interesting to further study this compound to verify whether this could be a realization of the mixed phases found in this paper.

These effects highlight the importance of SOC when considering the magnetic order of the iron pnictides and additionally provide an interesting avenue of future research into the impact of anisotropic magnetic fluctuations on superconductivity.

\begin{acknowledgments}
The authors are grateful to A. E. B{\"o}hmer, W. R. Meier, J. Kang, A. Kreisel, M. N. Gastiasoro, D. D. Scherer, and M. Sch{\"u}tt for valuable discussions. M.H.C. and R.M.F. were supported by the U.S. Department of Energy, Office of Science, Basic Energy
Sciences, under Award number DE-SC0012336. B.M.A. acknowledges financial
support from a Lundbeckfond fellowship (Grant No. A9318). P.P.O. acknowledges
support from Iowa State University Startup Funds. 
\end{acknowledgments}

\appendix

\section{Feynman diagrams}\label{app:diagrams}

\begin{figure}[b]
\centering
\includegraphics[width=1.0\columnwidth]{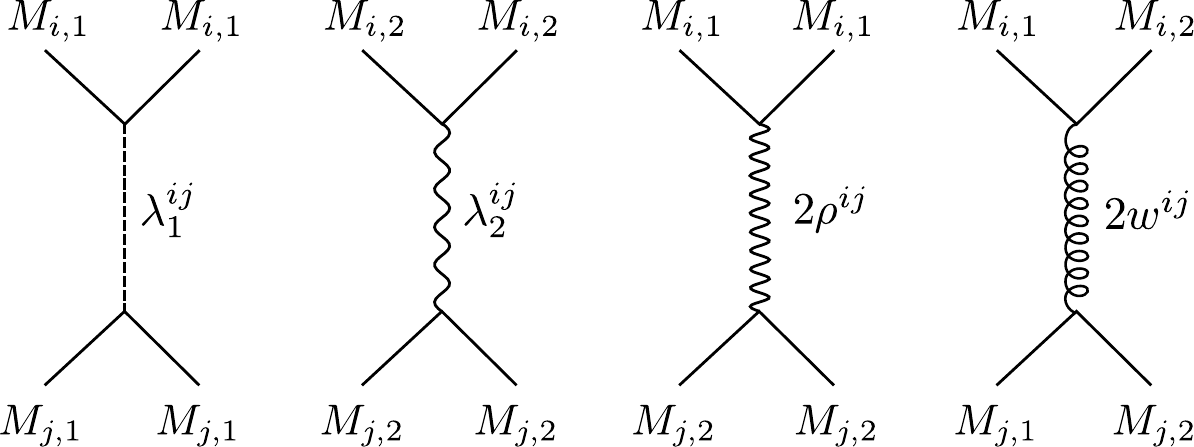}
\caption{\label{fig:app:bare_quartic_vertices} Illustration of bare quartic vertices. The legs of the diagrams correspond to the various spin components.}
\end{figure}

\begin{figure}
\centering
\includegraphics[width=1.0\columnwidth]{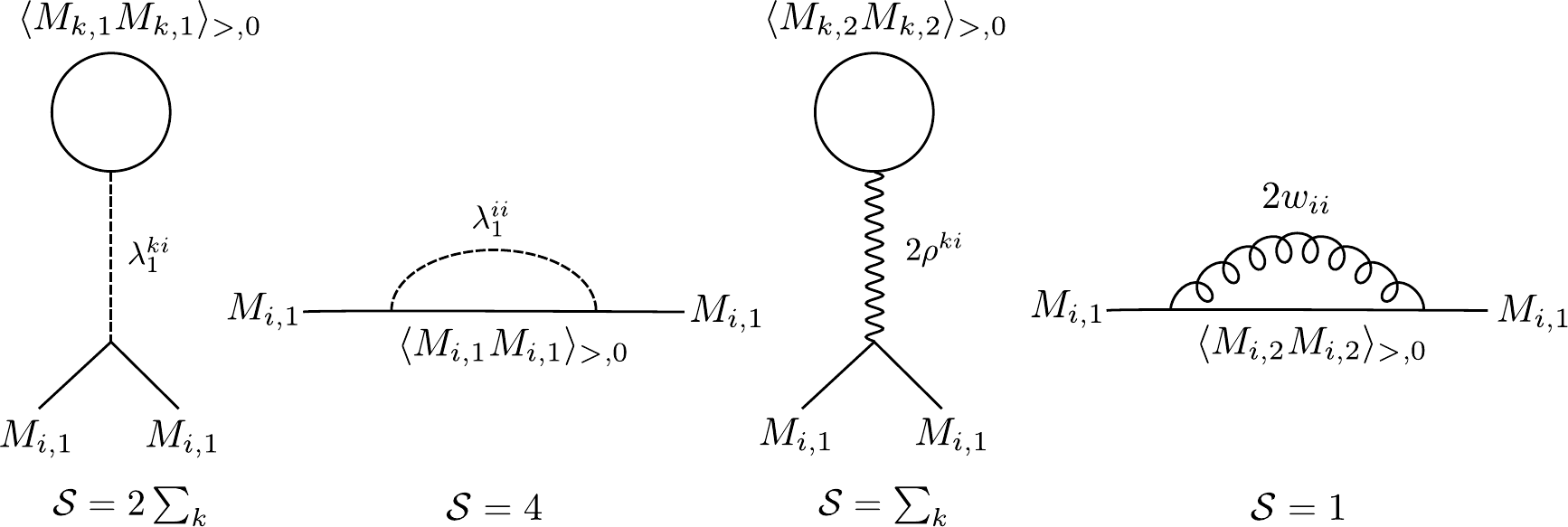}
\caption{\label{fig:app:renormalized_propagator} Diagrams contributing to the renormalization of the magnetic propagators. Here $\mathcal{S}$ denotes the symmetry factor, with $\sum_k$ arising from the closed bosonic loop. Note that these are just the diagrams contributing to the flow of $\tilde{r}_{i,1}$, a similar set with $1 \leftrightarrow 2$ exists which yields the flow of $\tilde{r}_{i,2}$.}
\end{figure}

In this appendix we present the relevant Feynman diagrams for deriving the RG flow equations in the presence of SOC. The bare quartic vertices are depicted in Fig.~\ref{fig:app:bare_quartic_vertices}.
In Fig.~\ref{fig:app:renormalized_propagator} we show the diagrams contributing to the renormalization of the propagator $\langle M_{i,1} M_{i,1} \rangle$ thus providing the flow of $\tilde{r}_{i,1}$.
Note that a set of similar diagrams exist, which contribute to the renormalization of $\langle M_{i,2} M_{i,2} \rangle$. To obtain these one can simply switch $1 \leftrightarrow 2$ in Fig.~\ref{fig:app:renormalized_propagator}. The symmetry factor of the diagrams is accounted for by the factor $\mathcal{S}$, which can contain a sum over components of the order parameters, $\sum_k$, arising from the closed bosonic loop. In the isotropic case this simply yields a factor of $N$ with $N$ being the number of components of the order parameter ($N=3$ here). In the evaluation of the diagrams in Fig.~\ref{fig:app:renormalized_propagator} the following integral enters:
\begin{eqnarray}
	\frac{\Omega_4}{(2\pi)^4}\int_{\Lambda e^{-\ell}}^{\Lambda}\mathrm{d}q \frac{q^{3}}{\tilde{r}_{i,A}+q^2} &=& 
	\frac{\Omega_4}{(2\pi)^4}\Lambda^{2}\frac{\ell}{1 + \tilde{r}_{i,A}}\,.\label{eq:prop_integral}
\end{eqnarray}
Here we used the fact that $d+z=4$ such that the integrals become four-dimensional. $\Omega_4=\frac{2\pi^{2}}{\Gamma(2)}$ is the area of a $3$-sphere, and we have rescaled the quartic coefficients by a factor of $\Omega_d /(2\pi)^4$. In the evaluation we assumed $\ell$ to be infinitesimal such that the integration is over a shell of thickness $\Lambda \ell$, within which the variation of $q$ can be neglected. Note that the factor of $\Lambda^{2}$ in Eq.~(\ref{eq:prop_integral}) serves to ensure that a similar rescaling can be carried out for the terms in the flow equations not involving a momentum integration, such as the term originating from Eq.~(\ref{eq:rescaling}).

\begin{figure}[H]
\centering
\includegraphics[width=\columnwidth]{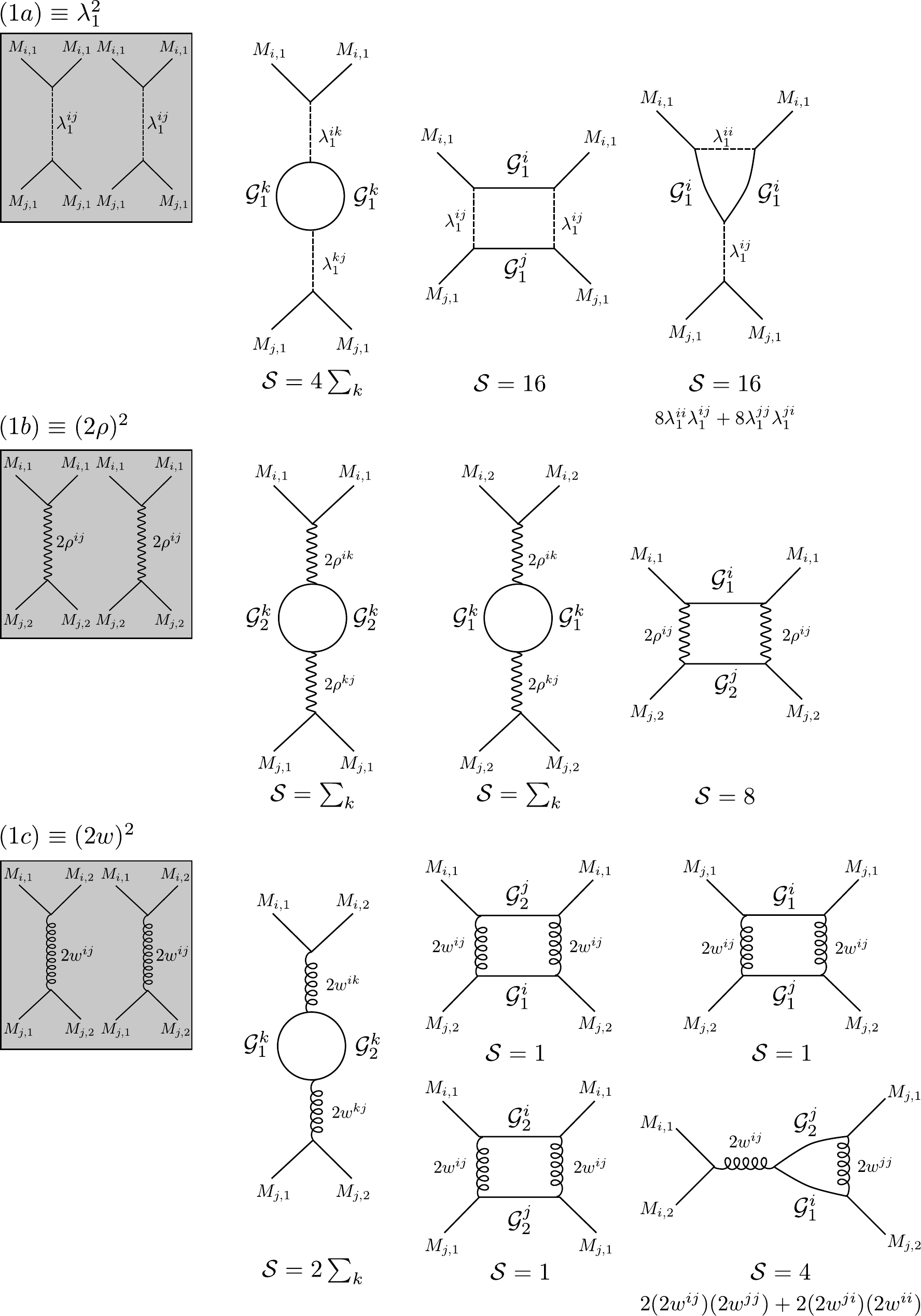}
\caption{\label{fig:app:first_class} Diagrams containing contributions from $\lambda_1^2$, $(2\rho)^2$ and $(2w)^2$. $\mathcal{S}$ denotes the symmetry factor of the respective diagram. The $\sum_k$ appearing in some symmetry factors is a consequence of the closed bosonic loop. Note that the diagrams with contributions from $\lambda_2^2$ can be obtained from $(1a)$ by switching $1 \leftrightarrow 2$.}
\end{figure}

\begin{figure}[H]
\centering
\includegraphics[width=\columnwidth]{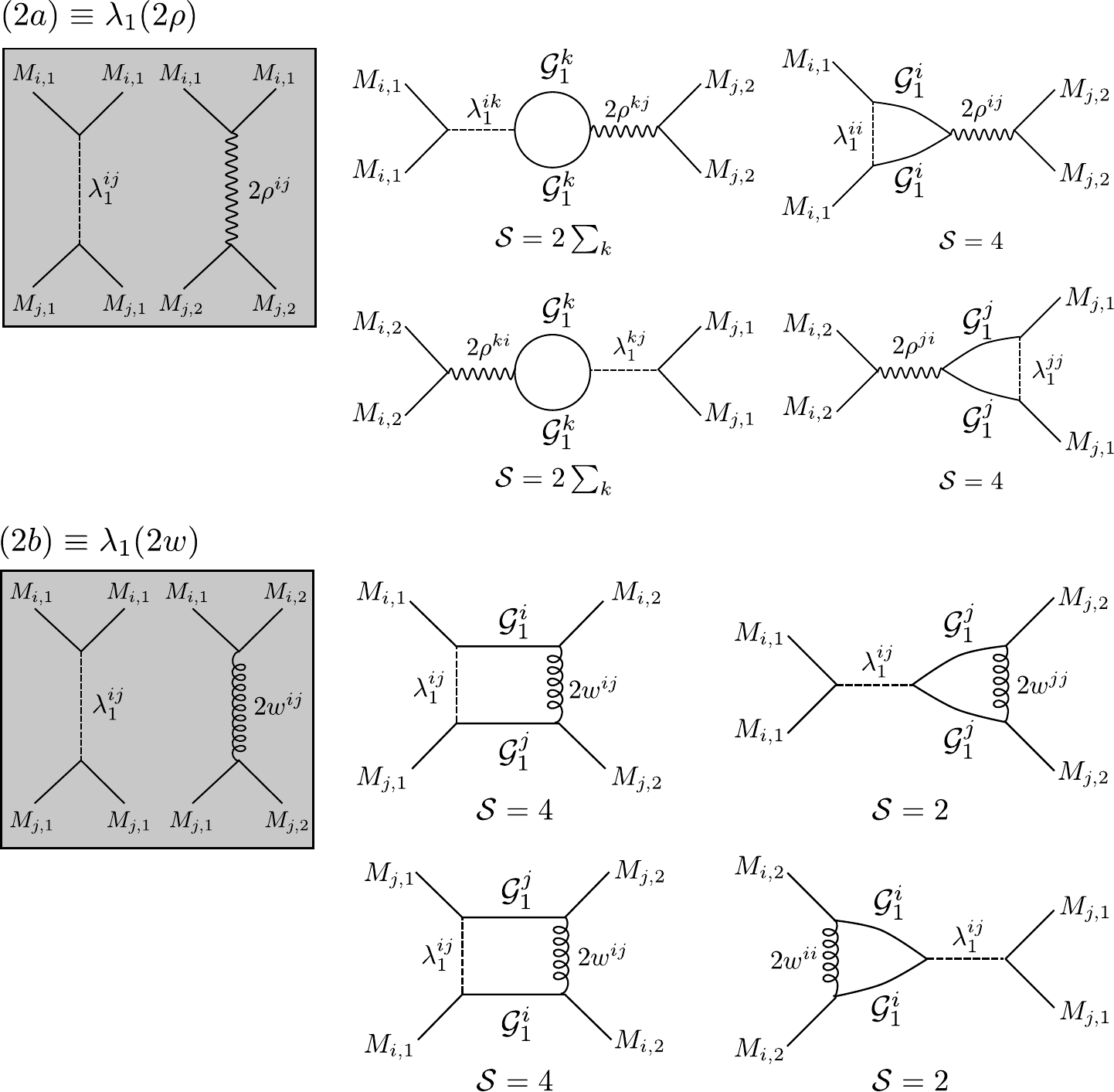}
\caption{\label{fig:app:second_class_1} Diagrams containing contributions from $\lambda_1 (2\rho)$ and $\lambda_1 (2w)$. As in Fig.~\ref{fig:app:first_class} the diagrams containing $\lambda_2$ can be obtained from the above by switching $1 \leftrightarrow 2$.}
\end{figure}

\begin{figure}[H]
\centering
\includegraphics[width=\columnwidth]{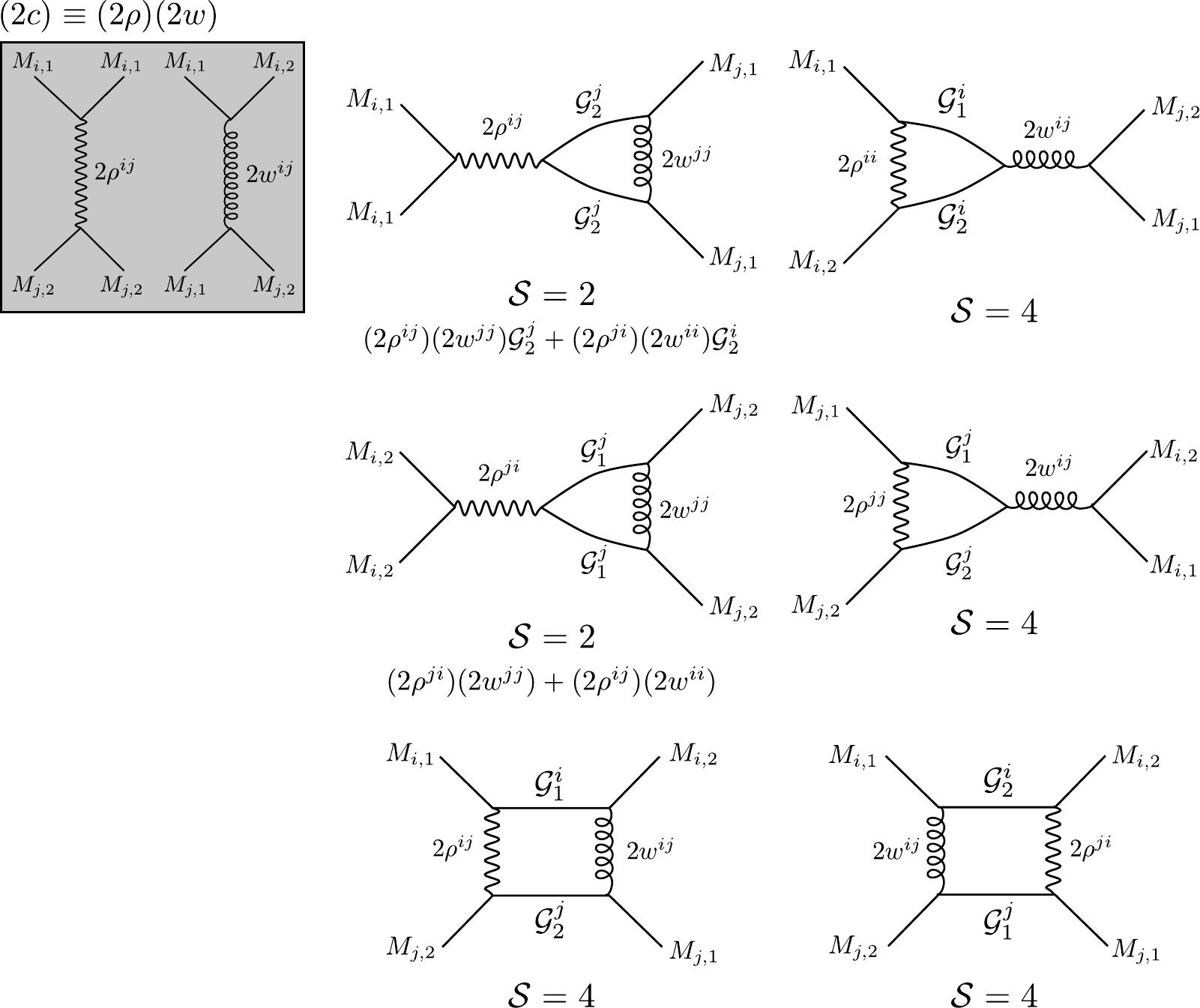}
\caption{\label{fig:app:second_class_2} Diagrams with contributions proportional to $(2\rho)(2w)$.}
\end{figure}

The diagrams contributing to the renormalization of the quartic coefficients can be divided into two classes: One obtained from combining identical bare vertices (Fig.~\ref{fig:app:first_class}), thus yielding contributions such as $\lambda_1^2$, and one obtained from combining distinct vertices (Figs.~\ref{fig:app:second_class_1} and \ref{fig:app:second_class_2}), yielding contributions like $(2\rho)(2w)$. We note that the term $-\left\langle \mathcal{S}_{\rm int} \right\rangle^2_{>,0}$ serve to cancel disconnected diagrams. The Green functions used in these diagrams can be denoted by a single index since
\begin{eqnarray}
	\mathcal{G}_1^{ij} &=& \begin{pmatrix}
		\frac{1}{r_0 + q^2 + \alpha_1} & 0 & 0 \\
		0 & \frac{1}{r_0 + q^2 + \alpha_2} & 0 \\
		0 & 0 & \frac{1}{r_0 + q^2 + \alpha_3}
	\end{pmatrix} \nonumber \\ 
	&=& \frac{\delta^{ij}}{r_0 + q^2 + \delta_{ix}\alpha_1+\delta_{iy}\alpha_2+\delta_{iz}\alpha_3} \\
	\mathcal{G}_2^{ij} &=& \begin{pmatrix}
		\frac{1}{r_0 + q^2 + \alpha_2} & 0 & 0 \\
		0 & \frac{1}{r_0 + q^2 + \alpha_1} & 0 \\
		0 & 0 & \frac{1}{r_0 + q^2 + \alpha_3}
	\end{pmatrix} \nonumber \\
	&=& \frac{\delta^{ij}}{r_0 + q^2 + \delta_{ix}\alpha_2 + \delta_{iy}\alpha_1 + \delta_{i3}\alpha_3}\,.
\end{eqnarray}
For the diagrams in the first class we omitted diagrams containing $\lambda_2^2$. These can be obtained by switching $1 \leftrightarrow 2$ in the diagrams $(1a)$ in Fig.~\ref{fig:app:first_class}. Similarly, we omitted the diagrams with $\lambda_2$ in Fig.~\ref{fig:app:second_class_1}; as above, these can be obtained by switching $1 \leftrightarrow 2$.
In evaluating the diagrams in Figs.~\ref{fig:app:first_class}--\ref{fig:app:second_class_2} we make use of the integral
\begin{eqnarray}
	 \frac{\Omega_4}{(2\pi)^4}\int_{\Lambda e^{-\ell}}^{\Lambda} && \mathrm{d}q \frac{q^{3}}{\left(\tilde{r}_{i,A} + q^2\right)\left(\tilde{r}_{j,B}+ q^2\right)} \nonumber \\  &&= 
	 \frac{\Omega_d}{(2\pi)^4} \frac{1}{\left(\tilde{r}_{i,1} + 1\right)\left(\tilde{r}_{j,1} + 1\right)}\,. \label{eq:second_class_integral}
\end{eqnarray}

\end{document}